\documentclass[a4paper,12pt,english]{article}

\usepackage{amsfonts,bm,amssymb,euscript,array,babel,cite}
\usepackage{amsmath,amsthm}
\usepackage{caption}

\makeatletter

\newcommand{\dd}{\mathrm{d}}

\newcommand{\w}{\wedge}
\newcommand{\bbm}{\left(\begin{matrix}}
\newcommand{\ebm}{\end{matrix}\right)}
\newcommand{\beq}{\begin{eqnarray}}
\newcommand{\eeq}{\end{eqnarray}}
\makeatother

\newcommand{\sfrac}[2]{{\textstyle\frac{#1}{#2}}}

\newcommand{\be}{\begin{equation}}
\newcommand{\ee}{\end{equation}}

\newcommand{\beqa}{\begin{eqnarray}}
\newcommand{\eeqa}{\end{eqnarray}} 
\def\nn{\nonumber} \def \bea{\begin{eqnarray}} \def\eea{\end{eqnarray}}

\newcommand{\barr}{\begin{array}}
\newcommand{\earr}{\end{array}}
\numberwithin{equation}{section}
\def\a{\alpha}  
  
 \def\d{\delta} 
    \def\k{\kappa}
   
 \def\o{\omega}

\def\mc{\mathcal}

\def\R{{\mathbb R}}  \def\N{{\mathbb N}}

 \def\one{\mbox{1 \kern-.59em {\rm l}}}

\def\bit{\begin{itemize}} \def\eit{\end{itemize}}

\def\({\left(} \def\){\right)}

 \allowdisplaybreaks[3]

\begin{document}

\makeatother

%%%%%%%%%%%%%%%%%%%%%%%%%%%%%%%%%%%%%%%%%%%%%%%%%%%%%%%%%%%%%%%5%%%%%%%%%%%%%%%%%%%%%%%%%%

\parindent=0cm

\renewcommand{\title}[1]{\vspace{10mm}\noindent{\Large{\bf

#1}}\vspace{8mm}} \newcommand{\authors}[1]{\noindent{\large

#1}\vspace{5mm}} \newcommand{\address}[1]{{\itshape #1\vspace{2mm}}}

%\thispagestyle{hepth}

%%%% --- TITLE PAGE --- %%%%
\begin{titlepage}
\begin{flushright}
 ITP-UH-03/14 
\end{flushright}
\begin{center}
\title{ {\LARGE Phase space quantization, noncommutativity and the gravitational field}}

\vskip 3mm

\authors{\normalsize Athanasios {Chatzistavrakidis}\footnote{thanasis@itp.uni-hannover.de}}

\vskip 3mm

\address{  {}Institut f\"ur Theoretische Physik,  
Leibniz Universit\"at Hannover, \\ Appelstra{\ss}e 2, 30167 Hannover, Germany }

\bigskip

\vskip 1.4cm

%%%% --- ABSTRACT --- %%%%
\textbf{Abstract}
\vskip 3mm

\begin{minipage}{14cm}%

In this paper we study the structure of the phase space in noncommutative geometry 
in the presence of a nontrivial frame. Our basic assumptions are that 
the underlying space is a symplectic and parallelizable manifold. 
Furthermore, we assume the validity of the Leibniz rule and the 
Jacobi identities. We consider noncommutative spaces due to the 
quantization of the symplectic structure and determine the momentum 
operators that guarantee a set of canonical commutation relations, 
appropriately extended to include the nontrivial frame. We stress the 
important role of left vs. right acting operators and of symplectic 
duality. 
This enables 
us to write down the form of the full phase space algebra on these noncommutative spaces, both in the noncompact and in the compact case.
 We test our results against the class of 
4D and 6D symplectic nilmanifolds, thus presenting a large set of 
nontrivial examples that realize the general formalism.

\end{minipage}

\end{center}

\end{titlepage}

\tableofcontents
\newpage

\section{Introduction} 

The most challenging conceptual problem of modern theoretical physics is the lack of a complete understanding 
of the physics of phenomena related to the fundamental constants $G_{N}$ and $\hbar$. These are quantum gravitational phenomena that become important 
near the Planck scale, defined by the mass scale $m_P=\sqrt{\frac{\hbar c}{G_N}}$, or by the length scale $l_P=\sqrt{\frac{\hbar G_N}{c^3}}$.

Quantum field theory describes physical processes where both c and $\hbar$ are important. In this sense, it is a unifying framework for 
special relativity and quantum mechanics. Its success is unquestionable, since it successfully incorporates three of the four 
fundamental interactions, the electromagnetic, weak and strong ones, in a particular unifying theory, the standard model. 
The latter has achieved unprecedented agreement with experimental data and a unique corroboration of its merit as a valid 
theory, at least up to energy scales of 1 TeV. 

Incorporating gravitational interactions in a unifying scheme with the rest of the forces is a notoriously difficult problem. 
String theory is at present the only framework where this is possible in a mathematically consistent way. The fundamental degrees of freedom in string theory are one-dimensional objects that do not propagate on a predetermined spacetime continuum, 
but instead they determine the geometry of spacetime. Indeed, the concept of spacetime and its dynamics is a derived or emergent concept 
that arises from the quantization of the two-dimensional non-linear sigma model that models the propagation of a string world sheet. 

The fact that strings are extended objects means that they cannot be associated to points in spacetime. This indicates that string geometry should have no points, unlike classical differential geometry. Pointless geometries are best accommodated in the 
mathematical framework of noncommutative geometry \cite{Connes:1994yd,Madorebook,Szabo:2001kg}. In such geometries, the spacetime coordinates become noncommuting operators and 
therefore a single point cannot by definition be resolved in any thought experiment, much like points in the phase space of quantum mechanics. This close relation between the geometry probed by a string and noncommutative spacetime geometries has been confirmed through the many connections that were established between string theory and noncommutative geometry \cite{Connes:1997cr,Seiberg:1999vs} (see also Ref. \cite{Blumenhagen:2014sba} for a review of more recent progress 
and the related literature).
In this sense, these two frameworks are close collaborators regarding questions of quantum gravitational phenomena.

String theory and noncommutative geometry introduce their respective scales, namely the string length $l_s$ in the first case and the 
length scale $l_{\text{NC}}$ where the classical description of spacetime is lost in the second. Equivalently one can think in terms of the string slope parameter $\a'$, which is equal to the square of the string length, or a noncommutativity scale $l^2_{\text{NC}}$ that 
appears in the commutator of coordinate operators in a noncommutative algebra. The above arguments suggest that there should exist a relation $\a'\sim l^2_{\text{NC}}$ among the two.
However, the generally accepted statement that at extremely small distances the classical notion of spacetime breaks down and it has to be 
replaced by some notion of quantum or fuzzy spacetime can be supported with arguments that are independent of string theory too 
\cite{dfr,Madorebook}. 

The physical motivation for the present work is to derive some lessons on the interplay between quantum mechanics and gravity through 
noncommutative geometry. In order to do so, one has to understand how the scales $G_{N}$ and $\hbar$ come together in a 
noncommutative algebra. 
This can be understood in the context of a noncommutative phase space, whose algebraic structure is in general
\bea 
[\hat x^a,\hat x^b]&=&il_{\text{NC}}^2\theta^{ab}(\hat x^c)~,\nn\\
{[}\hat x^a,\hat p_i]&=&i\hbar e^a_{\ i}(\hat x^c)=i\hbar \d^a_{\ i}+im_{\text{NC}}f^a_{\ ib}\hat x^b+\mc O(\hat x^2)~,\nn\\
{[}\hat p_i,\hat p_j]&=&i\sfrac{\hbar^2}{l_{\text{NC}}^2}F_{ij}~,\nn
\eea
where all the undefined quantities will be explained in detail below. What we point out here are the scales that 
appear in the algebra{\footnote{We set $c=1$, since we are not interested in the corresponding physics here.}}. 
As already discussed, the length scale $l_{\text{NC}}$ should be related to a small fundamental length, 
for example the Planck length $l_{\text{NC}}^2\sim G_{N}\hbar$ (or the string 
length, if one wishes to relate the two scales). Moreover, we denoted as $m_{\text{NC}}$ the combination 
$\hbar/l_{\text{NC}}$ (recall that $c=1$) and this should be related to the Planck mass, $m_{\text{NC}}^2\sim \hbar/G_N$.
It will become clear in the following that turning off the gravitational field the commutation relation between 
positions and momenta becomes the canonical commutation relation of quantum mechanics, and the momenta commute, as long 
as there are no magnetic sources in the problem. Thus turning off gravity we get noncommutative quantum 
mechanics and furthermore standard quantum mechanics in the limit $l_{\text{NC}}\to 0$.

The reason that we consider the full phase space instead of just the commutator of coordinates should be clear by the fact that 
quantum mechanics appears as a limiting case. Indeed, quantum mechanics teaches us that the phase space is an essential concept in the understanding of the underlying physics, a fact that is sometimes overlooked in applications of noncommutative geometry in high-energy physics.
It is reasonable to expect that phase space plays an equally important role in quantum 
gravity{\footnote{This statement can be made more precise once a dynamical theory for quantum fields in phase space
that incorporates the gravitational field is established. Although we do not directly address this problem in the present paper, 
we discuss a 
possible way to achieve this goal in the discussion section.}}. It should be 
mentioned that this was already emphasized long ago by Madore \cite{Madorebook}, who examined the role and properties of momenta 
in noncommutative geometry within the noncommutative frame formalism. 
More recent developments in this framework include Refs. \cite{Buric:2006di,Buric:2007hb,Buric:2011dd}. The importance of 
noncommutative phase space in physical problems, 
such as quantum particles in strong magnetic fields, was also emphasized in Refs. \cite{Duval:2000xr,Nair:2000ii,Morariu:2001dv,Horvathy:2002wc,Horvathy:2010wv}.

In this paper we are interested in examining the algebraic properties of the phase space when it is quantized in the presence of 
a nontrivial frame. The main physical reason to do this is that the frame is associated to the gravitational field. 
Therefore we expect to get some first lessons for physical problems that involve the behavior of quantum particles in 
the presence of gravity. These are situations that generalize the cases studied in Refs. 
\cite{Duval:2000xr,Nair:2000ii,Morariu:2001dv,Horvathy:2002wc,Horvathy:2010wv,Zhang:2001xs,Karabali:2001te,Karabali:2002im}, relevant for 
physical problems such as quantum particles moving in electromagnetic fields or the quantum Hall effect. 
On the other hand, although the noncommutativity of phase space in quantum mechanics was based on experimental facts, 
there is no experimental result yet that points to the phase space we describe here. However there are good 
conceptual reasons to consider it, as described above, and moreover one could hope for some basic experimental support of the 
general framework
by experiments such as 
the Fermilab Holometer \cite{Chou:2009zz,Berchera:2013lsa}, which is designed to test proposals associated to 
the quantization of spacetime.

In order to carry out the above task, one basic assumption we make is that the spaces we investigate admit a symplectic structure. It is well-known that symplectic manifolds 
have tractable quantization properties, either via deformation quantization \cite{Kontsevich:1997vb} or Weyl quantization \cite{Madore:2000en}. Moreover, 
we assume that the space is parallelizable so that a globally well-defined frame exists on it. Nontrivial symplectic, parallelizable and curved 
manifolds exist and we are going to provide a class of examples, the symplectic nilmanifolds. 

At this point it is useful to recall that in classical mechanics in d-dimensional flat space, 
the Hamiltonian formalism includes a set of coordinates $x^a,a=1,\dots,\text{d}$ and momenta $p_a$, building up a 
2d-dimensional phase space in d dimensions. This phase space has the structure of a symplectic manifold with symplectic 
structure,
\be
\o=\d^b_a\dd x^a\wedge \dd p_b~,
\ee
where summation is implied. $x^a$ and $p_a$ are the canonical coordinates with differentials 
(1-forms)
$\dd x^a$ and $\dd p_a$. The corresponding dual derivations are $\partial_{x^a}=\partial/\partial x^a$ and $\partial_{p_a}=\partial/
\partial p_a$. The symplectic structure defines a Poisson bracket, given as 
\be 
\{f,g\}=\d^b_a
(\partial_{x^b}f\partial_{p_a}g-\partial_{p_a}f\partial_{x^b}g)~.
\ee
In particular,
\be
\{x^a,p_b\}=\d^a_b~.
\ee

On the other hand, in quantum mechanics, where $x^a$ and $p_a$ become hermitian operators, the structure of $\dd x^a$, $\dd p_a$, 
$\partial_{x^a}$ and $\partial_{p_a}$ exhibits a degree of redundancy in the following sense. Since the wave function $\Psi (x)$ in 
the coordinate representation is a function of the positions only and not of both 
positions and momenta $\Psi(x,p)$, 
the quantum physics in the coordinate representation does not involve $\partial_{p_a}$ at all. Indeed, the 
canonical commutation relation (CCR)
\be
[\hat x^a,\hat p_b]=i\hbar\d^a_b~,
\ee
which can be thought of as the quantization of the Poisson bracket{\footnote{In the sense of Dirac, where the replacement 
$\{\cdot,\cdot\}\to \sfrac 1{i\hbar}[\cdot,\cdot]$ accounts for quantization. This simple relation should be treated cautiously, 
since it is not a sufficient and complete rule. For more details see for example the lecture notes \cite{weinstein}. }}, 
is represented on the Hilbert space by the operators
\be
\hat x^a\Psi= x^a\Psi, \quad \hat p_a\Psi=-i\hbar\d^b_a\partial_{x^b}\Psi~,
\ee
in accord with the Stone-von Neumann theorem.
In the dual picture of the momentum representation, where the wave function depends on the momenta $\Psi(p)$~,
the operators are represented as 
\be
\hat x^a\Psi= i\hbar\d^a_b\partial_{p_b}\Psi, \quad \hat p_a\Psi=p_a\Psi~.
\ee
Of course there is a continuum of intermediate mixed pictures but these are not 
particularly useful.
In any case, a simultaneous consideration of $\partial_{x^a}$ and $\partial_{p_a}$ is unnecessary. From a different point of view, 
employing the position representation, there is a commutative algebra of operators $\hat x^a$ and the momentum operators are included 
as outer derivations in the algebra and thus they do not belong to the algebra which the position operators generate. 

On the contrary to the latter statement, in noncommutative geometry the momenta do not necessarily correspond to outer derivations 
and they can as well be elements of the noncommutative algebra $\mc A$ generated by the position operators; namely they can 
also be inner. Thus, the starting point 
is the noncommutative but associative algebra $\mc A$ of coordinate operators and the momenta can be formally expressed in 
terms of the coordinate operators, $\hat p_a=\hat p_a(\hat x^b)$. This leads to a picture where the full phase space is associated to a 
noncommutative algebra where not only coordinates but also momenta do not necessarily commute among themselves. In the simplest case of 
noncommutative quantum mechanics in the absence of curvature, the CCRs are retained and they are 
supplemented by commutation relations among coordinates and among momenta separately. 
In the absence of sources and gravity, the momenta commute. However, this is not anymore true where sources are included or the gravitational
field is present.

In the present work we are interested in the case of the gravitational field, associated with a 
frame{\footnote{Throughout this paper, $i,j,\dots$ are flat (tangent space) indices, while 
 $a,b,\dots$ are curved (world) indices.} $e^i_{\ a}$. As already stated, we assume that the space is 
parallelizable and it admits 
a symplectic structure and that the symplectic 
2-form $\o$, as well as the corresponding symplectic 2-vector $\theta$, is constant in the basis of the globally well-defined frame. 
noncommutativity is introduced in the commutator of the position operators, setting it equal to the components of the 
symplectic structure in the curved basis,
\be
[\hat x^a,\hat x^b]=i\theta^{ab}(\hat x^c)~,
\ee
where from now on we set $l_{\text{NC}}=1$.
In this basis the parameters are not necessarily constant. However, we will see that 
there exists an interesting class of noncommutative spaces with curvature 
where they are constant in a chosen coordinate system.
In the presence of a nontrivial frame $e^i_{\ a}$ with inverse $e^a_{\ i}$, the appropriate commutation relations 
between momenta and coordinates are augmented to
\be
[\hat x^a,\hat p_i]=i\hbar e^a_{\ i}(\hat x^c)~,
\ee
where on the right hand side we encounter the noncommutative frame which is related to the gravitational field \cite{Buric:2006di}.  
In the following we will find the expression for the momenta 
and show how the full phase space algebra is determined. This algebra is required 
to satisfy the Jacobi identities. 
%This will give us the algebraic structure of the quantum phase space in the presence of a 
%general frame. 
Going one step further, we depart from the noncompact case and study the phase space algebra when a 
periodicity condition that compactifies the space is imposed. This is analogous to the condition that compactifies a d-plane to a d-torus. 
In such cases, the operators $\hat x^a$ turn out to be unphysical and the correct position operators are obtained with 
exponentiation of $\hat x^a$. 
Moreover, we test our results both in the noncompact and compact cases in a class of explicit examples,
the symplectic nilmanifolds in four and six dimensions. 

\section{Quantum mechanics and noncommutativity}

\subsection{Phase space of the noncommutative plane and torus}

In standard quantum mechanics, phase space is noncommutative. This is just the statement that a nontrivial commutation relation 
between positions and momenta exists, the CCR,  which is the basis of the uncertainty principle. 
However, both the position space and the momentum space are commutative. The full phase space algebra is simply
\bea\label{qmps}
[\hat x^i,\hat x^j]=0~, \quad [\hat x^i,\hat p_j]=i\hbar\d^i_j~, \quad [\hat p_i,\hat p_j]=0~,
\eea 
where we use flat indices for both positions and momenta{\footnote{In the present case 
the classical globally well defined 1-forms are simply $e^i=\d^i_a\dd x^a=\dd(\d^i_ax^a)$ and therefore we can introduce 
flat coordinates $x^i=\d^i_ax^a$ and the corresponding quantum operators. Similarly, the dual vector fields are just $\theta_i=\d_i^a\partial_a$ and we can 
define $\hat p_i=\d_i^a\hat p_a$. Thus we can work fully in flat indices instead of cluttering with the tensors 
$\d^a_i$ and $\d^i_a$.}.
As discussed in the introduction, this algebra of operators can be represented in the position representation, where the eigenvalues 
of the Hermitian operators $\hat x^i$ are ordinary real numbers $x^i\in \R$ and the momentum operators are translations, namely partial derivatives with 
respect to $x^i$, or in the dual momentum representation where the roles are exchanged. As we already stressed above, the momenta are introduced as outer 
derivations in the algebra of position operators. In standard quantum mechanics the momentum operators cannot be 
inner derivations of the algebra.

On the other hand, one can consider the quantum mechanics of particles on a noncommutative space, as for example in 
Refs. \cite{Duval:2000xr,Nair:2000ct,Nair:2000ii,Morariu:2001dv,Horvathy:2002wc,Horvathy:2010wv}. The simplest possibility is a noncommutative plane in 
d dimensions or a noncommutative d-torus. Although these cases are well known, let us review the main steps and results in order to warm up for the more 
general cases that we will present in the rest of this paper.

Let us first consider the noncompact case of a noncommutative d-plane 
of even dimension. This is specified by a set of coordinate operators which 
satisfy a commutation relation of the form $[\hat x^i,\hat x^j]=i\theta^{ij}$. Here, $\theta^{ij}$ is the set of 
constant noncommutativity parameters. They can be identified with the components of a symplectic structure on the corresponding classical manifold, 
namely with a constant symplectic 
2-vector in the globally well-defined basis. The quantum mechanics of particles on this space is associated with the noncommutative phase space algebra
\be
[\hat x^i,\hat x^j]=i\theta^{ij}~, \quad [\hat x^i,\hat p_j]=i\hbar\d^i_j~, \quad [\hat p_i,\hat p_j]=0~,
\ee
which extends the standard algebra (\ref{qmps}){\footnote{We are interested in the case where no magnetic sources are 
present. Otherwise the commutator of the momenta is also nonvanishing and proportional to the magnetic field \cite{Morariu:2001dv}.}}. 
The prime question to address concerns the realization of this algebra, which cannot be the same as in standard quantum mechanics for 
obvious reasons. One way to think about this problem is to associate $\hat x^i$ to a set of Hermitian matrices from a matrix algebra 
$\mc{A}$ (in particular 
the Heisenberg algebra in d dimensions) and the momenta to inner derivations of the algebra $\mc{A}$, similarly to Ref. \cite{Gross:2000wc}. Although 
representing momenta with inner derivations was impossible in standard quantum mechanics, it is 
working perfectly in the noncommutative case. Inner derivations in matrix algebras act with the adjoint action and therefore 
we consider the following Ansatz for the momenta,
\be
\hat p_i=c_{ij}\text{ad} \hat x^j = c_{ij}[\hat x^j,\cdot]~,
\ee
where $\cdot$ is a placeholder for arbitrary elements of the algebra $\mc{A}$ and $c_{ij}$ are constants to be determined. 
First we consider the CCRs with the position operators. Acting on an arbitrary function $f\in\mc{A}$ they imply
\bea
i\hbar \d^j_if&=&[\hat x^j,\hat p_i]f= \hat x^jc_{ik}[\hat x^k,f]-c_{ik}[\hat x^k,\hat x^jf]=\nn\\
&=&c_{ik}\hat x^j[\hat x^k,f]-c_{ik}\hat x^j[\hat x^k,f]-c_{ik}[\hat x^k,\hat x^j]f=\nn\\
&=&-ic_{ik}\theta^{kj}f~,
\eea
where we used the fact that $c_{ij}$ are constant and we applied the Leibniz rule. Then
 we obtain
\be
c_{ij}=\hbar \d_i^k\o_{kj}=\hbar \o_{ij}~,
\ee
where $\o_{ij}$ are the constant components of the non-degenerate symplectic 2-form which satisfies the relation 
\be \theta^{ij}\o_{jk}=-\d^i_k \ee
 with the symplectic 2-vector components. 
Therefore the momenta are 
given as 
\be\label{momplane}
\hat p_i=\hbar\o_{ij}[\hat x^j,\cdot]~.
\ee
Their trivial commutation relation remains to be examined. We compute
\bea
0&=&[\hat p_i,\hat p_j]f=\hbar\o_{ik}[\hat x^k,\hbar\o_{jl}[\hat x^l,f]]-
\hbar\o_{jl}[\hat x^l,\hbar\o_{ik}[\hat x^k,f]]=\nn\\
&=&\hbar^2\o_{ik}\o_{jl}([\hat x^k,[\hat x^l,f]]-[\hat x^l,[\hat x^k,f]])=\nn\\
&=&\hbar^2\o_{ik}\o_{jl}[[\hat x^k,\hat x^l],f]=i\hbar^2\o_{ik}\o_{jl}[\theta^{kl},f]~,
\eea
which holds because of the constancy of $\theta^{kl}$. In this computation the only additional input that has to be used is the Jacobi identity in 
the algebra $\mc{A}$, which we assume is valid.

A different but equivalent point of view is to write the Ansatz for the momentum operators as 
\be 
\hat p_i=c_{ij}(\hat x^j+\hat y^j)~,
\ee
with some appropriate operators $\hat y^j$. Under the assumption that 
\be
[\hat x^i,\hat y^j]=0~,
\ee
the commutation relation with $\hat x^i$ fixes $c_{ij}$ to be $\hbar\o_{ij}$, as before.
Then the vanishing commutator among the momenta gives
\be
0=[\hat p_i,\hat p_j]=\hbar^2\o_{ik}\o_{jl}([\hat x^k,\hat x^l]+[\hat y^k,\hat y^l]) \quad \Rightarrow \quad
[\hat y^i,\hat y^j]=-[\hat x^i,\hat x^j]~.
\ee
This means that 
\be
[\hat y^i,\hat y^j]=-i\theta^{ij}~.
\ee
In other words, the momenta are realized with two mutually commuting copies of the algebra $\mc A$. Although this might seem puzzling, 
it is fully equivalent to the previous approach, as explained in Ref. \cite{Gross:2000wc}. The equivalence is established upon the identification
\be
\hat y^i=-\hat x^i_R~,
\ee
where $\hat x^i_R$ denotes the generators of the 
${\mc A}_R$ copy of the algebra $\mc A$ which act on states from the right, instead of acting from the left as 
implicitly assumed up to now. The right and left acting copies of $\mc A$ are mutually commuting indeed. 
Then the momenta become 
\be 
\hat p_i=\hbar\o_{ij}(\hat x^j_L-\hat x^j_R)~,
\ee
which is the same as the expression (\ref{momplane}). It should be appreciated that whenever a commutator 
action is encountered in noncommutative theories, there are two copies of the algebra $\mc A$ involved, the left and the right 
acting ones. We will see in the following that this distinction between left and right acting operators becomes 
crucial in more involved cases than the planar one.

The above considerations provide an understanding of the phase space algebra of noncommutative quantum mechanics 
and its realization in the 
absence of sources. The next step is to consider the compact case, which corresponds to a noncommutative torus. 
We follow the analysis of Ref. \cite{Morariu:2001dv} in order to illustrate this case. The starting point of the analysis is that 
the standard periodicity condition of a d-torus has to be imposed, i.e.
\be \label{torusshift}
\hat x^i \sim \hat x^i + 2\pi R^j\d_j^i~,
\ee
where $R_i$ are the d radii of the corresponding cycles, not necessarily equal for a rectangular torus.
The central observation is 
that due to this condition, the operators $\hat x^i$ are not single valued and therefore 
they are unphysical, i.e. they do not correspond to observables. The physical operators of positions on the noncommutative torus 
are obtained by exponentiation as
\be 
X^i=e^{ib^i_j\hat x^j}~,
\ee
where $b^i_j$ are constants to be determined. This is achieved by demanding the unitary operators $X^i$ to be globally well defined, i.e. 
to be invariant under the shift (\ref{torusshift}). The condition for this to happen is
\be
ib^i_j2\pi R^k\d^j_k=2\pi iN \d^i_k~,\quad N\in\N~,
\ee
which gives
\be
b^i_j=\sfrac N{R^j}\d^i_j~.
\ee
This results in the well-defined operators
\be\label{Xto}
X^i=e^{\sfrac {i\hat x^i}{R^i}}~,
\ee
where we set $N=1$ for simplicity.
Then the stage is set to write down the full phase space algebra for the noncommutative torus. The momenta are given as before 
and they are now outer derivations instead of inner, which was the noncompact case. The algebra is
\bea
X^iX^j&=&e^{-\sfrac{i\theta^{ij}}{R^iR^j}}X^jX^i~,\\
\hat p_iX^j&=&X^j(\hat p_i+\sfrac{\hbar}{R^j}\d^j_i)~,\label{toruspX}\\
{[}\hat p_i,\hat p_j]&=&0~,
\eea
in agreement with the approach of Ref. \cite{Connes:1997cr} on noncommutative tori.
Representations and quantum bundles over this algebra were discussed in detail in Refs. \cite{Morariu:2001dv,Brace:1998ku}. Here we are interested 
in the generalization of the above phase space algebras in the presence of a nontrivial frame.

\subsection{Left vs. right action and symplectic duality}

Before introducing curvature, it is useful to discuss further the left and right realizations of the noncommutative algebra $\mc A$. To this end 
we return to curved indices and we consider the algebra $\mc A$, generated by $\hat x^a$. 
The latter satisfy the relation $[\hat x^a,\hat x^b]=i\theta^{ab}$, with the parameters $\theta^{ab}$ being in the curved basis and therefore 
not necessarily constant. 
We denote as $\hat x^a_L$ the left acting 
position operators and as $\hat x^a_R$ the right acting ones, and ${\mc A}_L$ 
and ${\mc A}_R$ denote the corresponding algebras.
Clearly, $[\hat x^a_L,\hat x^b_L]=i\theta^{ab}$. It is also obvious that 
$
[\hat x^a_L,\hat x^b_R]=0
$ in full generality.
On the other hand, for $f\in {\mc A}$ we find
\be
[\hat x^a_R,\hat x^b_R]f=(\hat x^a_R\hat x^b_R-\hat x^b_R\hat x^a_R)f=f(\hat x^b\hat x^a-\hat x^a\hat x^b)=f(-i\theta^{ab})=-i\theta^{ab}f+i[\theta^{ab},f]~.
\ee
In the flat case, the last term on the right-hand side is zero and then $\hat x_R^a$ form a copy of the algebra $\mc A$, as in 
Section 2.1. This is not generally true when the curved components of the symplectic structure are not 
constant. In order to be able to proceed further, two additional minimal assumptions are due. First, we demand that 
$f\in {\mc A}_L$~, a very mild assumption which is made 
anyway in all similar approaches. Secondly, in order to obtain $[\theta^{ab},f]=0$ with general $\theta^{ab}$, we assume that 
these components depend on the right acting operators $\hat x^a_R$ if they are not constant. Then due to the commutativity between left and right acting 
operators, we get
\be
[\hat x^a_R,\hat x^b_R]f=-i\theta^{ab}f~.
\ee
This means that the full set of relations is
\be
[\hat x_L^a,\hat x^b_L]=i\theta^{ab}~, \quad [\hat x^a_L,\hat x^b_R]=0~, \quad [\hat x^a_R,\hat x^b_R]=-i\theta^{ab}~.
\ee
The different sign of the left and right commutators makes manifest the symplectic duality among the two sets.
The concept of symplectic dual is very simple. Given a symplectic manifold with symplectic structure $\o$, its symplectic dual 
is again a symplectic manifold based on the same underlying manifold and an opposite symplectic structure $-\o$ \cite{weinstein}. This symplectic duality is 
elegantly realized in the context that we examine here.

Let us now take a look at the momentum operators too. In the flat case, 
the CCR $[\hat x^j,\hat p_i]=i\hbar \d^j_i$ holds as it is for the left 
acting position operators. However, it is a straightforward calculation to show that it holds the same for the right acting 
ones too, 
i.e.
\be
[\hat x^j_R,\hat p_i]=i\hbar\d^j_i~.
\ee
This obviates the need for right acting momentum operators. This should be expected from the fact that these operators already 
involve both the left and right acting position operators. 
On the other hand,
 in the curved case we will show that although
 $[\hat x_L^a,\hat p_i]=i\hbar e^a_{\ i}$,
in the most interesting cases it holds that 
\be 
[\hat x^a_R,\hat p_i]=i\hbar e^a_{\ i}+\hbar \o_{cb}[\hat x^a_R,e^c_{\ i}](\hat x^b_L-\hat x^b_R)~.
\ee 
This shows that left and right operators are liberated and they play asymmetric roles in the formalism. 
 Finally, it should be clear that had we focused on the momentum representation, the roles of 
momentum and position operators would have been fully exchanged, as in standard quantum mechanics. In this paper we work 
on the position representation.

\section{Quantized phase spaces and curvature}

\subsection{Introduction of curvature}

In the previous section we reviewed the phase spaces of standard quantum mechanics, noncommutative planar quantum mechanics and 
noncommutative toroidal quantum mechanics. In all cases the space is flat and there is no sign of the gravitational field. 
This is evident by the fact that only flat indices appear. Essentially, the flat tensor $\d^i_a$ is implicitly present in all 
the formulas, through the relation $\hat x^i=\d^i_a\hat x^a$ that holds in the flat case.

An elegant way  to introduce the gravitational field is through the vielbein and the frame formalism, as for example in the 
treatment of gravity as a gauge theory (see e.g. the textbook \cite{Ramond:1981pw} for a concise presentation). 
There one substitutes{\footnote{Up to now we did not bother about the horizontal position of the indices, since only the flat tensor appeared where there is 
no difference when the inverse or the transpose is taken.
From now on the position is important 
so let us explain how we denote these tensors. We do not use an explicit notation for the inverse or the transpose. It should be clear from the index 
structure. The inverse of $e^i_{\ a}$ is $e^{a}_{\ i}$ and the transpose is $e_a^{\ i}$. The inverse transpose is $e_i^{\ a}$. 
Also we often refrain from explicitly writing the $x$-dependence of all these tensors.}}}
\be
\d^i_{\ a}\quad \to \quad  e^i_{\ a}( x)~.
\ee 
This substitution indicates that the CCRs are augmented to a set of 
\underline{e}xtended CCRs (ECCRs) 
\be\label{eccr}
[\hat x^a,\hat p_i]=i\hbar  e^a_{\ i}(\hat x^b)~.
\ee
Such an approach was advocated and followed 
in Refs. \cite{Buric:2006di,Buric:2007hb,Buric:2011dd} and many aspects 
of our approach are similar, although not identical.
In Eq. (\ref{eccr}), $ e^a_{\ i}$ is the (inverse of the) noncommutative vielbein, obtained by the commutative one upon promoting 
the coordinates to operators, as in standard quantum mechanics. Note that in Section 2 we emphasized the role of left and right 
acting operators. Although we indicated the dependence of the frame as $\hat x^b$, it is not a priori clear which of the two 
sets of operators is to be taken. Our approach is to keep an open mind and let the consistency of the formalism decide. 
It will in fact turn out that the frame depends on the right acting set of operators $\hat x_R^b$.

Let us give a list of our assumptions, which we already mentioned in the introduction, with some additional details.
First we assume that the classical manifold admits a symplectic structure. In other words it
is endowed with a nondegenerate closed 2-form $\o$, $\dd \o=0,$  which is invertible. 
Second, we assume that the classical manifold is parallelizable and therefore it admits 
a set of globally well-defined 1-forms $e^i$ which serve as a basis of its cotangent bundle. 
In this basis the symplectic 2-form is assumed to have constant coefficients.
In flat space the $e^i$'s are just $\dd x^i$, but this is no longer true for an arbitrary, possibly curved, manifold. 
In general, they are related to an explicit coordinate basis by
\be
e^i=e^i_{ \ a}(x)\dd x^a~.
\ee
In geometric terms $e^i_{ \ a}(x)$ is the twist matrix which relates the two bases. It is invertible and its inverse is denoted as $e_{\ i}^a(x),$ as already stated; thus
\be
\dd x^a=e^a_{\ i}e^i~.
\ee
In gravitational language, it defines the gravitational field, relating flat and curved indices.
Its exterior derivative is a 2-form, which can be expanded in the basis of the cotangent bundle,
\be 
\dd e^i=-\sfrac 12 f^i_{\ jk}e^{jk}~,
\ee
where from now on we use the notation $e^{ij}=e^i\w e^j$. These are essentially the Maurer-Cartan equations.
Solving for the coefficients, we get
\be
f^i_{\ jk}=2e^a_{\ [j}e^b_{\ k]}\partial_be^i_{\ a}~.
\ee
The symplectic 2-form can be expanded in the basis 2-forms too. It has the form
\be
\o=\sfrac 12\o_{ij}e^{ij}~,
\ee
where $\o_{ij}$ is independent of $x^a$. 
It is useful to present its components in the curved basis as well. We compute
\be
\o=\sfrac 12 \o_{ij}e^i_{\ a}e^j_{\ b}\dd x^{ab} \quad \Rightarrow \quad \o_{ab}=\o_{ij}e^i_{\ [a}e^j_{\ b]}~.
\ee
The symplectic 2-vector has the form 
\be 
\theta=\sfrac 12\theta^{ij}\theta_i\wedge \theta_j~,
\ee 
where the $\theta_i$'s are the dual vectors to the basis 1-forms $e^i$. 
Its components are opposite to the components of the inverse of the symplectic 
2-form, namely
\be 
\theta^{ij}=-(\o^{-1})^{ij}~.
\ee 
Obviously, it holds that 
\be 
\theta^{ij}\o_{jk}=-\d^i_k~,
\ee 
and similarly for the curved indices.

Moreover, we assume the validity of the Leibniz rule and the Jacobi identities. The first means that for any three 
functions $f,g,h\in \mc A_L$
\be
[f,gh]=g[f,h]+[f,g]h~.
\ee
The Jacobi identities are
\be 
\text{Jac}(f,g,h):=[f,[g,h]]+[h,[f,g]]+[g,[h,f]]=0~.
\ee
These two assumptions are valid in the compact case too.
In fact we extend the above requirements to the full ${\mc A}_L\times {\mc A}_R$ 
algebra. Although our pool of configuration space
observables lies in ${\mc A}_L$, this is important because
$\hat x^a_R$ appear in the phase space algebra too.

Our interest is to construct the quantum mechanical phase spaces of noncommutative manifolds with a nontrivial vielbein, 
guided by Eq.~(\ref{eccr}). 
In this process, it is often needed to perform a Weyl ordering and we will explain when this is needed and how it is implemented.
We always assume that the quantization is performed along the symplectic structure of the manifold and therefore the 
commutator of position operators corresponds to the components of the symplectic 2-vector in the curved basis.
On the other hand,  we 
do not assume anything for the commutator among the momenta and instead we 
are going to derive it.
According to the above, the notation we use is 
\bea
[\hat x^a,\hat x^b]&=&i\theta^{ab}~,\label{crx} \\
{[}\hat x^a,\hat p_i]&=&i\hbar e_{\ i}^{a}~,\\
{[}\hat p_i,\hat p_j]&=&iF_{ij}~.
\eea
When we write $\hat x^a$ without a subscript, we implicitly mean the left acting 
operators. The right acting ones will always be indicated explicitly as 
$\hat x^a_R$, and their commutation relation 
with the momentum operators should be implemented in the above algebra.
This will be done explicitly in the following.
Finally, let us recall that this notation is appropriate for the noncompact cases. As we saw in the example of the noncommutative torus, the position operators 
have to be exponentiated in the compact case. We will denote these exponentiated operators as $X^a$, as we did in the previous section. 

\subsection{Momentum operators in the presence of a nontrivial frame}

Let us proceed a step further and determine the general properties of the position and momentum operators for an arbitrary 
symplectic parallelizable manifold in d dimensions. We begin with the noncompact case. 
We would like to 
determine the momenta which guarantee that the mixed commutation relation $[\hat x^a,\hat p_i]=e^a_{\ i}$ is satisfied. 
To this end we consider a similar Ansatz as for the case with trivial frame in Section 2,
\be 
\hat p_i=~: c_{ia}(\hat x^b_L,\hat x^b_R)(\hat x^a_L-\hat x^a_R):~,
\ee
with two notable differences than previously. First, we let the quantities $c_{ia}$, which have to be determined, to depend 
on $\hat x^a$. For the moment this dependence can be both on the left and right acting operators but we will see 
that a reduction of this dependence is necessary. The above Ansatz and the $\hat x^a$ dependence of $c_{ia}$ immediately introduces an 
ordering issue. This is similar to quantum mechanics, when 
products of position and momentum operators are encountered. The usual recipe is to introduce a Weyl ordering of the 
operators, denoted as $:\cdot:$, such as 
\be
:\hat x\hat p:=\sfrac 12(\hat x\hat p+\hat p\hat x)~,
\ee
and similarly for higher order ambiguities. Presently, the Weyl ordering means that{\footnote{If the $c_{ia}$'s are higher than 
linear order, then they are also normal ordered. We refrain from explicitly indicating this in our notation.}}
\be
\hat p_i=\sfrac 12 (c_{ia}\hat x_L^a+\hat x^a_L c_{ia}-c_{ia}\hat x^a_R-\hat x^a_R c_{ia})~.
\ee

Having specified the Ansatz, let us insert it in the ECCRs in order to determine the unknown functions $c_{ia}$:
\bea
i\hbar e^b_{\ i}=[\hat x^b,\hat p_i]&=&
\sfrac 12(\hat x^b_Lc_{ia}\hat x^a_L+\hat x^b_L\hat x^a_Lc_{ia}-\hat x^b_Lc_{ia}\hat x^a_R-
\hat x^b_L\hat x^a_Rc_{ia}-\nn\\&-&c_{ia}\hat x^a_L\hat x^b_L-\hat x^a_Lc_{ia}\hat x^b_L
+c_{ia}\hat x^a_R\hat x^b_L+\hat x^a_Rc_{ia}\hat x^b_L)~.
\eea 
In order to be able to solve this condition, an assumption on the dependence of $c_{ia}$ has to be made. 
If we assume that the dependence is on $\hat x^a_L$, the commutation relation $[\hat x^a_L,\hat x^b_R]=0$ may be used to 
obtain
\be
i\hbar e^b_{\ i}=\sfrac 12(\hat x^b_Lc_{ia}\hat x^a_L+\hat x^b_L\hat x^a_Lc_{ia}-2[\hat x^b_L,c_{ia}]\hat x^a_R
-c_{ia}\hat x^a_L\hat x^b_L-\hat x^a_Lc_{ia}\hat x^b_L)~.
\ee
This equation is in general not sufficient to determine $c_{ia}$. The situation 
is greatly improved by the alternative choice of $c_{ia}$ depending on $\hat x^a_R$. In this case we are led to{\footnote{Had we made the 
additional assumption that $[c_{ia}(\hat x^c_L),\hat x^b_L]=0$ in the previous case, we would have arrived at the same expression. However, this route would not 
lead in general to consistent results.}}
\be
i\hbar e^b_{\ i}=c_{ia}(\hat x^c_R)[\hat x^b_L,\hat x^a_L]=-ic_{ia}(\hat x^c_R)\theta^{ab}~,
\ee
which implies 
\be 
c_{ia}(\hat x^c_R)=\hbar e^b_{\ i}\o_{ba}~.
\ee
It is important to stress that this works only if we assume that the noncommutative frame also depends on the right acting set of operators, 
which is in full agreement with the discussion and assumptions of Section 2.2.
This was up to now not specified but it is forced on us by the computation itself and the consistency of the algebra.
Therefore we find that the momenta are given as
\be\label{momentagen}
\hat p_i=\hbar~:e^a_{\ i}\o_{ab}(\hat x_L^b-\hat x^b_R):=\hbar:e^a_{\ i}\o_{ab}[\hat x^b,\cdot]:~.
\ee
A simple consistency check shows that when the frame is trivialized the momenta of the noncommutative plane are recovered. 
It is observed that in the presence of a nontrivial frame the momenta are complicated operators. Due to the noncommutativity of 
the frame, it is difficult to proceed in a general evaluation of the momentum commutator $F_{ij}$ and to 
present a closed form for the conditions due to the Jacobi identities. These tasks are tractable once the 
frame is given and the symplectic form is known. This will be the case in the next section.

However, let us examine the momentum commutator under some assumptions, which 
will prove valid in the next section. First, let us assume that 
$\o_{ab}$, and therefore $\theta^{ab}$ too, are constant parameters{\footnote{This might seem strange, 
since they carry curved indices, but in Section 4 we will show that it 
is a relevant case.}}. Then the momenta are simplified to 
\be\label{momentasimp}
\hat p_i=\hbar\o_{ab}~:e^a_{\ i}(\hat x_L^b-\hat x^b_R):~.
\ee
We introduce the following notation
\bea
[e^a_{\ i},\hat x^b_L]&=&0~,\nn\\
{[}e^a_{\ i},\hat x^b_R]&=&K^{ab}_i~, \label{K}\\
{[}e^a_{\ i},e^b_{\ j}]&=&L^{ab}_{ij}~. \label{L}
\eea 
The first equation is due to the fact that the frame is $\hat x_R^a$-dependent. Moreover, it holds that $L^{ab}_{ij}=-L^{ba}_{ji}$ 
and we do not assume any other symmetry property for $L_{ij}^{ab}$ or $K^{ab}_{i}$~.
This allows us to compute the momentum commutator and find
the simple expression{\footnote{In order to reach this expression, we assume that $\o_{ac}\o_{bd}[e^d_{\ j},K^{cb}_i]=
\o_{ac}\o_{bd}[\hat x^b_R,L^{cd}_{ij}]=0$~, which can be checked retrospectively in specific cases.}}
\bea\label{pcrgen}
[\hat p_i,\hat p_j]=
\hbar^2\o_{ac}\o_{bd}\biggl(L^{cd}_{ij}(\hat x^a_L-\hat x^a_R)(\hat x^b_L-\hat x^b_R)
 -2K^{cb}_{[i}e^d_{\ j]}(\hat x^a_L-\hat x^a_R)\biggl)~.
\eea 
It is clear from the structure that the right hand side contains linear and quadratic terms in $\hat p_i$ and possibly constant terms too. 
We express this as 
\be\label{quadratic}
[\hat p_i,\hat p_j]=M_{ij}+N_{ij}^{\ k}\hat p_k+P_{ij}^{kl}\hat p_k\hat p_l~,
\ee
where $M_{ij},N_{ij}^{\ k},P_{ij}^{kl}$ are parameters antisymmetric in their 
lower indices, while $P^{kl}_{ij}$ is symmetric in its upper indices.
We should stress that a quadratic algebra for the momenta was proven by Madore to be the only consistent 
choice for matrix algebras \cite{Madorebook}. Comparing the last two expressions, 
we determine the coefficients to be
\bea \label{P}
P_{ij}^{kl}&=&e^k_{\ c}e^l_{\ d}L^{cd}_{[ij]}~,\\ \label{N}
N_{ij}^{\ k}&=&\hbar\o_{bd}e^k_{\ c}\biggl(2K^{cb}_{[i}e^d_{\ j]}+
P_{ij}^{ml}(K^{cb}_le^d_{\ m}+K^{db}_{(m}e^c_{\ l)})\biggl)~,\\ \label{M}
M_{ij}&=&-\sfrac{\hbar^2}4\o_{ac}\o_{bd}P_{ij}^{kl}K^{ca}_kK^{db}_l~.
\eea
Having determined the general form of $F_{ij}$, the last remaining piece is 
to examine the commutators of $\hat x^a_R$. These commute with $\hat x^a_L$ and 
they satisfy the opposite algebra to them (symplectic duality). The only unknown commutator is the one with the 
momentum operators, which we now compute and we find
\be \label{rightcom}
[\hat x^a_R,\hat p_i]=i\hbar e^a_{\ i}-\hbar\o_{bc}K^{ba}_{i}(\hat x^c_L-\hat x^c_{R})=i\hbar e^a_{\ i}-e^k_{\ b}K^{ba}_i\hat p_k~.
\ee  
Therefore, if the frame is known then the full 
phase space algebra is uniquely determined.

Before proceeding, let us comment on the compact case too. 
This proceeds along similar lines that led us from the d-plane to the d-torus, with some additional input due to the 
$\hat x^a_R$-dependence of the frame. 
A periodicity condition is
imposed, but this time there is a difference between the right acting and the left acting operators. 
The key to understand which conditions to impose for each set is to appreciate that $\hat x^a_L$ 
are to be associated to positions (property of e.g. a quantum mechanical particle), while $\hat x^a_R$ are to be associated 
with the frame or gravitational field and therefore to coordinates (property of space itself). This distinction between 
position and coordinate suggests that the conditions that compactify the space should be applied to the right acting 
operators, as
\be \label{nilperio}
\hat  x^a_R \sim \hat x^a_R+2\pi R^i\tau^a_{\ i}(\hat x^b_R)~,
\ee
with some appropriate $\hat x^a_R$-dependent tensor $\tau^a_{\ i}~$ that has to be specified. This tensor is related to 
$e^a_{\ i}$ but it is not the same as that, as it will 
become obvious in the next section. 
On the other hand, the positions, which are associated to the operators $\hat x^a_L$, should be single valued when we return on 
the same point after traversing a cycle of the compact space. For nontrivial elliptic fibrations, where the cycles are of toroidal 
nature, this means that the periodicity condition on the left acting operators should be analogous to Eq. (\ref{torusshift}), namely
\be 
\hat x^a_L\sim \hat x^a_L+2\pi R^i\d^a_{\ i}~.
\ee 
This is the correct condition because single valuedness has to be guaranteed along the cycle that is traversed and not along 
other directions too, since in this process the positions in the other directions might have changed due to the nontrivial fibration structure. Then the position operators are 
obtained by exponentiation as
 \be
 \label{Xnil}
 X^a=e^{\sfrac{i\hat x^a}{R^a}}~.
 \ee 
 The momenta are now outer, as in the toroidal case, as well as the operators $\hat x^a_R$ which do not correspond to observables and they do not necessarily have to be 
 exponentiated.
The phase space algebra in the compact case turns out to be
\bea \label{compgen1}
X^aX^b&=&e^{-\sfrac{i\theta^{ab}}{R^aR^b}}X^bX^a~,\\ \label{compgen2}
\hat p_iX^a&=&X^a(\hat p_i+\sfrac {\hbar}{R^a}e^a_{\ i})~,
\eea 
and the rest of the commutators remain the same as in the noncompact case.
Similar considerations appeared in Refs. \cite{Chatzistavrakidis:2012qj,Chatzistavrakidis:2012yp}, although in a 
different and less general context and without reference to symplectic structures. In particular, 
in those papers the commutator of either the positions or the momenta is set to zero, which is not 
a consistent choice in the present context, while Jacobi anomalies appear, which is at odds with our assumptions here.

\section{Quantized phase space of symplectic nilmanifolds}

The main general assumptions that we made are the existence of symplectic structure and parallelizability, as well as the Leibniz rule and 
the Jacobi identity. Having so far worked on the 
general case, it is now time to examine whether these assumptions include any nontrivial examples. We already know the trivial ones, 
which are the d-plane in the noncompact case and the d-torus in the compact case. On the other hand, spheres are completely excluded. 
The only symplectic sphere, $S^2$, is not parallelizable and the only parallelizable spheres, $S^3$ and $S^7$, are not symplectic{\footnote{
Spheres can be quantized with different methods, such as the ones used in Refs. \cite{Madore:1991bw,Ramgoolam:2001zx,Ramgoolam:2002wb,Hasebe:2010vp}.}}.

An interesting class of nontrivial symplectic and parallelizable manifolds is provided by group manifolds based on nilpotent 
Lie algebras and the associated compact nilmanifolds.
Therefore we would like to apply the above formalism to this class of spaces. We are going to work in dimensions 4 and 6. 
This is a choice 
based on the following reasons. First, symplectic manifolds are always even-dimensional{\footnote{This does not mean that odd-dimensional 
nilmanifolds cannot be quantized using their symplectic leaves. In Ref. \cite{Rieffel} a deformation quantization of the 3-dimensional 
Heisenberg nilmanifold is presented (see also Ref. \cite{Lowe:2003qy}).}}. Second, there are no 
nilmanifolds in 2D, apart from the trivial one of the 2-torus. Dimension 4 contains only two nontrivial cases of nilmanifolds 
\cite{Patera:1976ud,goze} and they are both symplectic. Symplectic nilmanifolds in 6D are also fully classified \cite{goze}.
They number 26 cases (or, more precisely, classes) and they may be read 
off either from Chapter 8, Section III of the book \cite{goze} or from the table at the end of Ref. \cite{sixmanifolds}. In this section we construct the quantized 
phase space of these 
spaces both in the noncompact and compact cases.

\subsection{Step classification of 4D and 6D nilmanifolds}

Nilmanifolds were introduced as simple examples of manifolds which admit symplectic structure 
but not K\"ahler structure{\footnote{The only 
K\"ahler nilmanifold is the d-torus \cite{kahlernil}. Moreover, nilmanifolds that do not admit a symplectic structure exist too \cite{sixmanifolds}.}}. From another point of 
view, they may be described as nontrivial generalizations of the torus, in the sense that any nilmanifold is an iterated twisted 
 fibration of toroidal fibers over toroidal bases. Employing this point of view, the globally defined basis of the 
cotangent bundle of a nilmanifold can be obtained in an elegant way by twisting the corresponding one for a flat torus. 
In particular, consider a nilpotent Lie algebra with structure constants $f^{a}_{\ bc}$. In accord with commonly used notation 
we present such an algebra as a d-tuple $(ab,cd,...,yz)$, whence its structure constants are $f^1_{\ ab},f^2_{\ cd},.\dots,f^
\text{d}_{\ yz}$.
Then we form the $(1,1)$ tensor 
\be 
F=\sfrac 12\mu_{(ab)} f^c_{\ ab}x^b\dd x^a\wedge\partial_c~,
\ee
parametrized by constants $\mu_{(ab)}$, where the indices between parentheses 
are not summed and there is no symmetry property that provides a
 direct relation between $\mu_{(ab)}$ and $\mu_{(ba)}$.
Exponentiating this tensor and acting with it on the basis 1-forms $\d^i_{\ a}\dd x^a$ of the d-torus we get
\be\label{Faction}
e^i=\d^i_{\ a}e^F\dd x^a~.
\ee
The action is performed with the standard interior product between vectors and forms. 
Since we are going to work with nilmanifolds up to dimension six, we use an expanded expression for the frame up to terms which are 
nonvanishing in such cases, i.e.
\bea\label{framegen}
e^i_{\ a}&=&\d^i_{\ a}+\sfrac 12\kappa_{(ab)}f^i_{\ ab}x^b+\sfrac 18\kappa_{(bcad)}f^{i}_{\ bc}f^{b}_{\ ad}x^cx^d+\sfrac 1{48}\kappa_{(bcpdaq)}f^i_{\ bc}f^b_{\ pd}f^p_{\ aq}x^cx^dx^q+ \nn\\
&+&\sfrac 1{384}\kappa_{(bcpdrqas)}f^i_{\ bc}f^b_{\ pd}f^p_{\ rq}f^r_{\ as}x^cx^dx^qx^s~.
\eea 
Note that this expression is more general than a simple expansion of Eq. (\ref{Faction}), since the constants $\kappa_{(\dots)}$ 
in front of each term are now not fixed by lower step terms. No symmetry properties for these constants are assumed.
The amount of non-vanishing terms on the right hand side is on a par with the nilpotency step of the underlying Lie 
algebra. For step 1 only the first term is there, which agrees with the fact that a step 1 nilmanifold is a torus. 
For step 2 we get the first two terms, since in this case it holds that 
\be\label{step2}
f^{i}_{\ bc}f^{b}_{\ ad}=0~,
\ee
even without summation in the index $b$,
by the definition of step 2.
For step 3, Eq. (\ref{step2}) is violated but it holds that 
\be\label{step3}
f^i_{\ bc}f^b_{\ pd}f^p_{\ aq}=0~,
\ee
again without summation in the repeated indices.
Similarly, for step 4, Eqs. (\ref{step2}) and (\ref{step3}) are violated but the following relation holds
\be 
f^i_{\ bc}f^b_{\ pd}f^p_{\ rq}f^r_{\ as}=0~.
\ee
Finally, the step 5 cases violate all the above relations. Step 6 nilmanifolds do not exist in six dimensions by definition. 
This is the reason that we stopped the expansion of the general formula to these five terms.

The basis 1-vectors can be determined in the same way. 
Moreover, the inverse $e^a_{\ i}$ of the frame $e^i_{\ a}$ satisfies 
\be
e^{i}_{\ a}e^a_{\ j}=\d^{i}_j \quad \text{and}  \quad e_a^{\ i}e_i^{\ b}=\d_a^b~,
\ee
and it is given by the analogous expanded formula
\bea\label{invframegen}
e^a_{\ i}&=&\d^a_{\ i}-\sfrac 12\lambda_{(ib)}f^a_{\ ib}x^b+\sfrac 18 \lambda_{(cbid)}f^{a}_{\ cb}f^{c}_{\ id}x^bx^d
-\sfrac 1{48}\lambda_{(cbpdiq)}f^a_{\ cb}f^c_{\ pd}f^p_{\ iq}x^bx^dx^q
+\nn\\
&+&\sfrac 1{384}\lambda_{(cbpdrqis)}f^a_{\ cb}f^c_{\ pd}f^p_{\ rq}f^r_{\ is}x^bx^dx^qx^s~,
\eea 
where 
\bea
\lambda_{(ib)}&=&\kappa_{(ib)}~,\nn\\
\lambda_{(cbid)}&=&-~\k_{(cbid)}+2\k_{(cb)}\k_{(id)}~, \nn\\
\lambda_{(cbpdiq)}&=&\k_{(cbpdiq)}-3\k_{(pdiq)}\k_{(cb)}+6\k_{(cb)}\k_{(pd)}\k_{(iq)}~,\nn\\
\lambda_{(cbpdrqis)}&=&-~\k_{(cbpdrqis)}+4(\k_{(pdrqis)}\k_{(cb)}+\k_{(cbpdrq)}\k_{(is)})+6\k_{(cbpd)}\k_{(rqis)}-\nn\\
&&-~12(\k_{(cbpq)}\k_{(rq)}\k_{(is)}+\k_{(rqis)}\k_{(pq)}\k_{(cb)})+24\k_{(cb)}\k_{(pd)}\k_{(rq)}\k_{(is)}~.\nn
\eea

Let us turn our attention to the parameters $\kappa_{(\dots)}$ that were introduced. 
Clearly, they are not arbitrary since they are constrained by the 
Maurer-Cartan equations. It is easy to determine this constraint if we 
focus on the step 2 case, where 
$$
e^i_{\ a}=\d^i_{\ a}+\sfrac 12\kappa_{(ab)}f^i_{\ ab}x^b~.
$$
We compute
\bea 
-\sfrac 12 f^i_{\ jk}e^{jk}&=&\dd e^i=\dd(e^i_{\ a}\dd x^a)=\partial_be^i_{\ a}\dd x^{ba}=e^b_{\ j}e^a_{\ k}\partial_be^i_{\ a}e^{jk}\nn\\
&\Rightarrow&f^i_{\ jk}=e^a_{\ j}e^b_{\ k}(\partial_be^i_{\ a}-\partial_ae^i_{\ b})\nn\\
&\Rightarrow& {\kappa_{(ab)}+\kappa_{(ba)}=2}~.\label{kappas}
\eea 
The most symmetric choice would be $\kappa_{(ab)}=\kappa_{(ba)}=1$, which was used for 
example in Ref. \cite{Chatzistavrakidis:2012qj}.  However, this is not the choice we make in the present paper 
and there is a very good reason for this, related to the symplectic structures 
on the nilmanifolds. We return to this immediately after we discuss these structures.

The classification of nilpotent (but not solvable) Lie algebras in four dimensions can be found in Ref. \cite{Patera:1976ud}. There are only 
three cases to consider. The first is the 4-torus, which is step 1 and symplectic. As we already mentioned, 
it is a rather degenerate case,  
in the sense that unlike all 
the other nilmanifolds, the torus is a K\"ahler manifold. Moreover, it is a flat space. Its symplectic 2-form can be chosen to be 
$\o=e^{12}+e^{34}~,$ with $e^i=\d^i_a\dd x^a$~. We will not discuss it further since it was already 
discussed in Section 2 in any dimension. The other two cases are given by $(0,0,0,12)$ and $(0,0,42,12)$~. The first one is a 
toroidal extension of the Heisenberg algebra in dimension three and it has nilpotency step 2. The second one has nilpotency step 3,
since it clearly contains the non-zero second order commutator of algebra generators $[T_2,[[T_1,T_2]]=-T_3$, or equivalently the non-vanishing 
quantity $f^3_{\ 42}f^{4}_{\ 12}$, but no third order non-vanishing commutator. Both cases admit a non-degenerate symplectic structure.
All these are summarized in the following table:
\begin{center}
\boxed{
\begin{tabular}{ccc}
 \underline{Class} & \underline{Step} & \underline{Symplectic form} 
 \\[4pt]
(0,0,0,12)  & 2 & $e^{14}+e^{23}$\\[4pt]
(0,0,42,12) & 3 & $e^{14}+e^{23}$ 
\end{tabular}}\captionof{table}{Nontrivial symplectic nilmanifolds in 4D.}
\end{center}

We move on to 6D, 
where there are 26 classes of symplectic nilmanifolds. We would like to sub-classify them according to 
their nilpotency step. 
As before, the 6-torus is the only one step 1 nilmanifold.
In the following four tables we present the step 2,3,4 and 5 6D symplectic nilmanifolds along with their symplectic 
2-form, the latter taken from the single table of Ref. \cite{sixmanifolds}.

\begin{center}
\boxed{
\begin{tabular}{cc}
 \underline{Class}  & \underline{Symplectic form}
 \\[4pt]
(0,0,0,0,0,12) & $e^{16}+e^{23}+e^{45}$\\[4pt]
(0,0,0,0,13+42,14+23)  &  $e^{16}+e^{25}+e^{34}$ \\[4pt]
(0,0,0,0,12,13)  &  $e^{16}+e^{25}+e^{34}$\\[4pt]
(0,0,0,0,12,34)  & $e^{15}+e^{36}+e^{24}$\\[4pt]
(0,0,0,0,12,14+23) & $e^{13}+e^{26}+e^{45}$\\[4pt]
(0,0,0,12,13,23)  & $e^{15}+e^{24}+e^{36}$
\end{tabular}}\captionof{table}{Step 2 symplectic nilmanifolds in 6D.}
\end{center}

\begin{center}
\boxed{
\begin{tabular}{cc}
 \underline{Class}  & \underline{Symplectic form}
 \\[4pt]
(0,0,0,0,12,14+25)  & $e^{13}+e^{26}+e^{45}$\\[4pt]
(0,0,0,0,12,15)  &  $e^{16}+e^{25}+e^{34}$ \\[4pt]
(0,0,0,12,14+23,13+42)  &  $e^{15}+2e^{26}+e^{34}$\\[4pt]
(0,0,0,12,14,13+42)  & $e^{15}+e^{26}+e^{34}$\\[4pt]
(0,0,0,12,14,23+24)  & $e^{16}-e^{34}+e^{25}$\\[4pt]
(0,0,0,12,13,14)  & $e^{16}+e^{24}+e^{35}$\\[4pt]
(0,0,0,12,13,24) & $e^{26}+e^{14}+e^{35}$\\[4pt]
(0,0,0,12,13,14+23)  & $e^{16}-2e^{34}-e^{25}$
\end{tabular}}\captionof{table}{Step 3 symplectic nilmanifolds in 6D.}
\end{center}

\begin{center}
\boxed{
\begin{tabular}{ccc}
 \underline{Class}  & \underline{Symplectic form}
 \\[4pt]
(0,0,0,12,14-23,15+34)  & $e^{16}+e^{35}+e^{24}$\\[4pt]
(0,0,0,12,14,15)  &  $e^{13}+e^{26}-e^{45}$ \\[4pt]
(0,0,0,12,14,15+24) &  $e^{13}+e^{26}-e^{45}$\\[4pt]
(0,0,0,12,14,15+23+24) &  $e^{13}+e^{26}-e^{45}$\\[4pt]
(0,0,0,12,14,23+15) &  $e^{13}+e^{26}-e^{45}$\\[4pt]
(0,0,12,13,23,14) &  $e^{15}+e^{24}+e^{34}-e^{26}$\\[4pt]
(0,0,12,13,23,14-25) &  $e^{15}+e^{24}-e^{35}+e^{16}$\\[4pt]
(0,0,12,13,23,14+25) &  $e^{15}+e^{24}+e^{35}+e^{16}$
\end{tabular}}\captionof{table}{Step 4 symplectic nilmanifolds in 6D.}
\end{center}

\begin{center}
\boxed{
\begin{tabular}{ccc}
 \underline{Class}  & \underline{Symplectic form}
 \\[4pt]
(0,0,12,13,14,15) & $e^{16}+e^{34}-e^{25}$\\[4pt]
(0,0,12,13,14,15+23)  & $e^{16}+e^{34}+e^{24}-e^{25}$\\[4pt]
(0,0,12,13,14+23,15+24)  & $e^{16}+2e^{34}-e^{25}$
\end{tabular}}\captionof{table}{Step 5 symplectic nilmanifolds in 6D.}
\end{center}

The above five tables contain 27 cases where we can apply the results of Section 3. It is 
interesting and welcome that a lot of diversity is exhibited, since there are indeed cases up to step 5 which do admit 
symplectic structure.

\subsection{Determining the phase space algebra}

It is obvious from the above tables that in the flat basis of $e^i$, the 
components of the symplectic 2-form and the corresponding 2-vector are 
constant. However, this is not in general true for the symplectic 2-form 
in an arbitrary curved basis. Let us discuss how and when the curved basis 
components $\o_{ab}$ and $\theta^{ab}$ can be constant as well. This is 
essentially the reason that we introduced the constants $\kappa_{(\dots)}$ 
previously, instead of making the symmetric choice for them.

We propose the following: 

\textbf{Proposition:} For 4D and 6D symplectic nilmanifolds whose 
symplectic structure has the same form as the corresponding torus, there exists a coordinate system, specified by a
choice of parameters $\kappa_{(\dots)}$~, such that the components $\o_{ab}$ ($\theta^{ab}$) of the 
symplectic 2-form (2-vector) in the curved basis are constant and equal to 
the flat components $\o_{ij}$ ($\theta^{ij}$).\\
\textbf{Proof:} It is a straightforward task to determine the form of $\kappa_{(\dots)}$ 
for each case separately and the corresponding frame $e^i_{\ a}$ that 
delivers the required result. 
The full list of results for the frame $e^i_{\ a}$ appears in the tables of the 
Appendix and proves the 
proposition.
$\square$

This proposition shows that all but four symplectic nilmanifolds satisfy 
the assumption that led to Eqs. (\ref{pcrgen}) and (\ref{quadratic}) 
with coefficients (\ref{P})-(\ref{M}), which are central 
for our purposes{\footnote{The four cases that are not covered  
are indicated accordingly in the tables of the Appendix and they are three step 4 cases and one step 5. 
They can also be found in Tables 4 and 5 as the ones whose symplectic form involves four summands.}}. This in turn means that the full phase space 
algebra can be determined in a closed form for all these cases. 
 To this end, the 
momentum commutator has to be fully determined and this will be the case  once we compute the quantities $K^{ab}_i$ and $L^{ab}_{ij}$, defined in Eqs. (\ref{K}) and (\ref{L}). 
This is possible 
because the frame is known explicitly in all cases. 

Let us work out in full detail the step 2 case. 
 We compute
\bea 
K^{ab}_i=-\sfrac i2 \kappa_{(ic)}f^a_{\ ic}\theta^{bc}
\eea 
and
\bea 
L^{ab}_{ij}=\sfrac i4 \kappa_{(ic)}\kappa_{(jd)}f^a_{\ ic}f^b_{\ jd}\theta^{dc}~.
\eea 
We directly obtain that
\be 
P_{ij}^{kl}=\sfrac i4 \kappa_{(ic)}\kappa_{(jd)}f^k_{\ [i\underline{c}}f^l_{\ j]d}\theta^{dc}~,
\ee 
due to Eq. (\ref{step2}), where the underlined index is excluded from 
the antisymmetrization. The same equation implies also that
\be 
P^{kl}_{ij}K_l^{ab}=0~.
\ee
Then the Eqs. (\ref{N}) and (\ref{M}) give
\be 
M_{ij}=0~, \quad N_{ij}^{\ k}=-i\hbar f^k_{\ ij}~,
\ee 
where Eq. (\ref{kappas}) was used.
Let us mention once more that the constants $\kappa_{(ia)}$ are known for each case, since they determine the frame components that guarantee the 
constancy of $\theta^{ab}$. 
This leads to the following phase space algebra in the step 2 case:
\be \label{s2alg}
{[\hat x^a,\hat x^b]=i\theta^{ab}, \quad [\hat x^a,\hat p_i]=i\hbar e^a_{\ i}, 
\quad [\hat p_i,\hat p_j]=-i\hbar f^k_{\ ij}\hat p_k+\sfrac i4 \kappa_{(ic)}\kappa_{(jd)}f^k_{\ [i\underline{c}}f^l_{\ j]d}\theta^{dc}\hat p_k\hat p_l}~.
\ee
Additionally, 
\be \label{s2algR}
[\hat x^a_R,\hat x^b_R]=-i\theta^{ab}~,\quad [\hat x^a_R,\hat p_i]=i\hbar
e^a_{\ i}+\sfrac i2\kappa_{(ic)}f^k_{\ ic}\theta^{ac}\hat p_k~. 
\ee 
The Jacobi identities have to be satisfied as well. 
Nontrivial ones include
\be 
\text{Jac}(\hat p_i,\hat p_j,\hat p_k)=0~,
\ee 
which is satisfied due to Eq. (\ref{step2}), and 
\be 
\text{Jac}(\hat p_i,\hat p_j,\hat x^a)=0 \quad \Rightarrow \quad [e^a_{\ i},\hat p_j]-
[e^a_{\ j},\hat p_i]=i\hbar f^a_{\ ij}-2P_{ij}^{kl}e^a_{\ (k}\hat p_{l)}~. 
\ee 
The latter is not obvious, but we have checked that it is satisfied in all step 2 cases, using the data of the Appendix. In the following, it will be examined 
in some explicit examples too, also for higher steps. The same holds for the Jacobi identities 
that involve $\hat x^a_R~$.

We point out that the resemblance to the planar case is exhausted in the constancy of the parameters $\theta^{ab}$. A comparison of 
Eqs. (\ref{s2alg}) and (\ref{s2algR}) with the corresponding ones of the planar case shows that the two algebras are very 
different.

\subsection{Remarks on the compact case}

As a final task, we examine the compact case.
This is 
captured by the general expressions (\ref{compgen1}) and (\ref{compgen2}), where the parameters 
$\theta^{ab}$ and $e^a_{\ i}$ are explicitly known for the cases at hand. 
Here we would like to include some additional remarks on the compactification of the group manifold, in order to make clear 
why the exponentiated coordinates are given by Eq. (\ref{Xnil}) in the nontoroidal case and they do not have to be 
modified from the toroidal ones of Eq. (\ref{Xto}).

Consider the simplest case of step 2 nilmanifolds, since 
the higher step cases are just direct generalizations of the procedure. 
Here we work on the classical manifold. The globally well defined 1-forms are 
\be 
e^i=(\d^i_a+\sfrac 12 \kappa_{(ab)}f^i_{\ ab}x^b)\dd x^a~.
\ee 
Consider shifting the coordinate $x^b\to x^b+2\pi \d^b_{\ i}R^i$~. Then 
\be 
e^i\to e^i_{\ a}\dd x'^a+\sfrac 12\kappa_{(ab)}f^i_{\ ab}\d^b_{\ i}2\pi R^i\dd x'^a~.
\ee 
Since the frame is global, it should remain invariant under this shift and 
this means that $x^a$ cannot remain the same. Indeed we observe that 
they have to change to
\be \label{xrshift}
x'^a=x^a-\sfrac 12\kappa_{(bi)}f^a_{\ bi}x^b2\pi R^i~.
\ee 
Then using again Eq. (\ref{step2}), we find that $e^i\to e^i$, as it should. Evidently, although the 1-forms $e^i$ are globally well 
defined, this is not true for the coordinates $x^a$. On the contrary, on a torus it is possible to define global coordinates 
besides global 1-forms. In other words, a shift around a cycle in the toroidal case leads back to the same point and therefore 
the position in every direction has to be exactly the same, which is achieved with exponentiation. In the nilmanifold case, 
a shift around a cycle associated with the topologically nontrivial fibration structure of the manifold leads to the same point in the 
direction of the cycle, but to a different point in the orthogonal directions, since the fibered tori have changed geometrically 
according to the nontrivial twist. This means that the position has to be single valued in the direction of the shift, but 
in the fiber directions this position will naturally change, exactly because the test particle never returned to the exact same position 
that it had before the shift in those directions. This result indicates that the single-valuedness along the shifted direction is taken care of by 
exponentiation, exactly as for the toroidal case. On the other hand, the change of position due to the twist is encoded in the 
commutation relation between the momenta (which are translation operators and as such they correspond to the generators of the shift) 
and the positions, as in Eq. (\ref{compgen2}).

As we explained in Section 3.2, the compactification of the nilmanifold in the noncommutative case requires a stronger periodicity 
condition on the operators $\hat x^a_R$ that provide the frame dependence. This is given by Eq. (\ref{nilperio}). 
Eq. (\ref{xrshift}) suggests that the overall periodic shift in the $\hat x^a_R$ is
\be 
\hat x^a_R\sim \hat x^a_R+2\pi R^i(\d^a_{\ i}-\sfrac 12\kappa_{(bi)}f^a_{\ bi}\hat x^b_R)~.
\ee  
This gives the dual noncommutative frame
\be \label{tau}
\tau^a_{\ i}=\d^a_{\ i}-\sfrac 12 \k_{(bi)}f^a_{\ bi}\hat x^b_R~
\ee 
for the step 2 case and we can see that this is different than $e^a_{\ i}$~, since $\kappa_{(ia)}$ has no symmetry property.
Its inverse is 
\be 
\tau^i_{\ a}=\d^i_{\ a}+\sfrac 12 \k_{(ba)}f^i_{\ ba}\hat x^b_R~.
\ee 
Although these considerations seem very different than the simple toroidal case, it should be reminded that in the case of 
the compactification of the plane one also encounters two different tori dual to each other, one associated to the 
periodicity condition (\ref{torusshift}) and one to the algebraic relation (\ref{toruspX}). The situation is analogous here, 
where there are two nilmanifolds, one associated to the inverse frame $e^a_{\ i}$ in Eq. (\ref{compgen2}) and one to the 
dual inverse frame $\tau^a_{\ i}$ of the periodicity condition (\ref{nilperio}). 

\section{Benchmark examples of quantized phase spaces}

\subsection{Dimension 4}

Let us now proceed and apply the general results in some benchmark cases 
from the tables of symplectic nilmanifolds, beginning with dimension four. 
There are only two cases, so we examine both of them. 

\paragraph{Step 2: (0,0,0,12)~.}

In the present case the only non-vanishing structure constant is the $f^4_{\ 12}=-f^4_{\ 21}=1$. The basis of the cotangent bundle is 
\be \label{frames2}
e^i=\dd x^i~, ~i=1,2,3~, \quad e^4=\dd x^4+x^2\dd x^1~.
\ee
In other words, $\kappa_{(12)}=2$ and $\kappa_{(21)}=0$. All the other parameters vanish.
It holds that
$$
\dd e^4=-e^{12}~,$$
which is essentially the Maurer-Cartan equation.
The symplectic structure in the natural basis is specified by the 2-form
\be 
\o=e^{14}+e^{23}
\ee
and it is easily confirmed that it is closed and non-degenerate.
Moreover, it is easily confirmed that 
$$e^{14}=\dd x^{14} \quad \text{and}\quad e^{23}=\dd x^{23}~.$$
The dual basis is spanned by the vectors 
\be 
\theta_1=\partial_1-x^2\partial_4~,\quad \theta_i=\partial_i~, ~i=2,3,4~.
\ee 
The symplectic 2-vector is{\footnote{Here $\theta_{14}=\theta_1\wedge \theta_4$ and it should not be confused with the constant component $\theta^{14}=1$~.}}
\be 
\theta=\theta_{14}+\theta_{23}=\partial_{14}+\partial_{23}~.
\ee 
Then we set
\be 
[\hat x^1,\hat x^4]=i \quad \text{and} \quad [\hat x^2,\hat x^3]=i~.
\ee 
 The 
fact that the only non-vanishing off-diagonal element of $e^i_{\ a}$ is the $e^4_{\ 1}= x^2$ means that 
for its inverse $e^a_{\ i}$ the corresponding component is equal to $-x^2$ and therefore the noncommutative frame will have $e^4_{\ 1}= -\hat x_R^2$~, according to 
the analysis of Section 3{\footnote{This can be directly derived from 
Eq. (\ref{invframegen}), or one can write the frame as a matrix and find its 
inverse, in particular 
$$
e^i_{\ a}=\begin{pmatrix} 1 &0&0&0 \\ 0&1&0&0 \\ 0&0&1&0 \\ 
x^2&0&0&1\end{pmatrix} \quad \Rightarrow \quad e^a_{\ i}=\begin{pmatrix} 1 &0&0&0 \\ 0&1&0&0 \\ 0&0&1&0 \\ 
-x^2&0&0&1\end{pmatrix}~,
$$
for the present example. }}.
This means that 
\be
[\hat x^4,\hat p_1]=-i\hbar \hat x^2_R~.
\ee
 The momenta are determined by Eq. (\ref{momentasimp}) 
and they turn out to be
\bea
\hat p_1&=&\hbar([\hat x^4,\cdot]+\hat x^2_R[\hat x^1,\cdot])~, \nn\\ \hat p_2&=&\hbar[\hat x^3,\cdot]~, \nn\\ \hat p_3&=&-\hbar[\hat x^2,\cdot]~, \nn\\ \hat p_4&=&-\hbar[\hat x^1,\cdot]~.\nn
\eea
We observe that no ordering is necessary in any of the four momentum operators. Computing the commutators of the momenta, we find 
that the only non-vanishing one is
\be 
[\hat p_1,\hat p_2]=-i\hbar\hat p_4~,
\ee
in accord with the general results of the previous section for the 
quantity $N_{ij}^{\ k}$~.
For the right acting operators we find the non-vanishing off-diagonal commutators
\be 
[\hat x^3_R,\hat p_1]=-i\hat p_4~,\quad [\hat x^4_R,\hat p_1]=-i\hbar x^2_R~.
\ee 
It is a straightforward task to show that all the Jacobi identities 
are satisfied identically.

Let us now consider the compact case. From Eq. (\ref{frames2}) we get that for the classical manifold
\be 
x^2\to x^2+2\pi R^2 \quad \Rightarrow \quad x^4\to x^4-2\pi R^2x^1~,
\ee
which means that the periodicity condition (\ref{nilperio}) in the noncommutative case 
involves the off-diagonal component
\be 
\tau^4_{\ 2}=-\hat x^1_R~.
\ee 
According to Eq. (\ref{Xnil}) the position operators are given by
\be 
X^a=e^{\sfrac i{R^a}\hat x^a}~, ~a=1,2,3,4~.\nn
\ee 
Then we can find the phase space algebra in the compact case, which is 
\bea 
X^1X^4&=&e^{-\sfrac i{R^1R^4}}X^4X^1~,\quad X^2X^3=e^{-\sfrac i{R^2R^3}}X^3X^2~, \nn\\
\hat p_1X^4&=&X^4(\hat p_1-\sfrac{\hbar}{R^4}\hat x^2_R)~,
\eea 
plus the diagonal mixed relations and the momentum commutators which were already written down above.

\paragraph{Step 3: (0,0,42,12)~.}

In this case the non-vanishing structure constants are $f^3_{\ 42}=f^4_{\ 12}=1$. The basis of the cotangent bundle is taken to be 
\be\label{frames3} 
e^i=\dd x^i~, ~i=1,2~,\quad e^3=\dd x^3-x^4\dd x^2-x^1x^2\dd x^2~,
  \quad e^4=\dd x^4+x^2\dd x^1~.
\ee
This time the only non-vanishing parameters are
 $\kappa_{(24)}=\kappa_{(12)}=2$ and $\kappa_{(4221)}=8$~.
The dual vectors are easily found,
\be 
\theta_1=\partial_1-x^2\partial_4~,\quad \theta_2=\partial_2+(x^4+x^2x^1)\partial_3~,
\quad \theta_i=\partial_i~, ~i=3,4~.
\ee 
The symplectic 2-form is
\be 
\o=e^{14}+e^{23}=\dd x^{14}+\dd x^{23}~,\nn
\ee
while the corresponding 2-vector is
\be 
\theta=\theta_{14}+\theta_{23}=\partial_{14}+\partial_{23}~,\nn
\ee 
and we set again
\be 
[\hat x^1,\hat x^4]=i \quad \text{and} \quad [\hat x^2,\hat x^3]=i~.
\ee 
 The only non-vanishing off-diagonal elements of $e^a_{\ i}$ are the 
$e^3_{\ 2}=-(x^4+x^2x^1)$ and $e^4_{\ 1}= x^2$; the noncommutative frame has
 $e^3_{\ 2}=\hat x^4_R+\hat x^1_R\hat x^2_R$ and $e^4_{\ 1}= -\hat x_R^2~$, which sets the 
commutation relations
\be
[\hat x^3,\hat p_2]=i\hbar(\hat x^4_R+\hat x^1_R\hat x^2_R)~,\quad [\hat x^4,\hat p_1]=-i\hbar
 \hat x^2_R~.
\ee
 The momenta are again found using Eq. (\ref{momentasimp}) 
and they turn out to be
\bea
\hat p_1&=&\hbar([\hat x^4,\cdot]+\hat x^2_R[\hat x^1,\cdot])~, \nn\\ \hat p_2&=&\hbar([\hat x^3,\cdot]-(\hat x^4_R+\hat x^1_R\hat x^2_R)[\hat x^2,\cdot])~, \nn\\ \hat p_3&=&
-\hbar[\hat x^2,\cdot]~, \nn\\ \hat p_4&=&-\hbar[\hat x^1,\cdot]~.\nn
\eea
The non-vanishing momentum commutators are
\be 
[\hat p_2,\hat p_4]=i\hbar \hat p_3~,\quad
[\hat p_1,\hat p_2]=-i\hbar\hat p_4~.
\ee
Once more, all the Jacobi identities are satisfied. The least trivial one is 
\bea  
\text{Jac}(\hat p_1,\hat p_2,\hat x^3)&=&[[\hat p_1,\hat p_2],\hat x^3]+
[[\hat x^3,\hat p_1],\hat p_2]+[[\hat p_2,\hat x^3],\hat p_1]=\nn\\
&=&0+0-i\hbar[\hat x^4_R+\hat x^1_R\hat x^2_R,\hat p_1]=\nn\\
&=&\hbar^2\hat x^2_R-\hbar^2\hat x^2_R=0~,
\eea  
where we used the fact that a direct computation gives $[\hat x_R^4,\hat p_1]=-i\hbar x^2_R~$. The rest of the commutators involving $\hat x^a_R$ are
\be 
[\hat x^3_R,\hat p_1]=-i\hat p_4~,\quad [\hat x^3_R,\hat p_2]=i\hbar( \hat x^4_R+ \hat x^1_R\hat x^2_R)+i\hat x^1_R\hat p_3~,\quad [\hat x^1_R,\hat p_2]=-i\hat p_3~,
\quad [\hat x^4_R,\hat p_2]=i\hat x^2_R\hat p_3~.\nn
\ee 
For the compact case, the frame (\ref{frames3}) suggests the nontrivial shifts 
\bea 
&& x^4 \to  x^4+2\pi R^4 \quad \Rightarrow \quad  x^3\to 
 x^3+2\pi R^4 x^2~,\nn\\
&& x^2\to  x^2+2\pi R^2\quad \Rightarrow \quad  x^4\to 
 x^4-2\pi R^2 x^1~,\nn\\
&& x^1\to  x^1+2\pi R^1 \quad \Rightarrow \quad x^3\to  x^3+\sfrac {2\pi R^1}2 ( x^2)^2~,\nn
\eea 
in the classical case.
This means that in the noncommutative case the off-diagonal $\tau^a_{\ i}$ components are
\be 
\tau^3_{\ 4}=\hat x^2_R~, \quad \tau^3_{\ 1}=\sfrac 12 (\hat x^2_R)^2~, \quad \tau^4_{\ 2}=-\hat x^1_R~.
\ee
Then it is straightforward to compute the relations among the $X^a$ and $\hat p_i$, 
which turn out to be (we write down only the nontrivial ones)
\bea 
X^1X^4&=&e^{-\sfrac{i}{R^1R^4}}X^4X^1~,\quad X^2X^3=e^{-\sfrac{i}{R^2R^3}}X^3X^2~,\nn\\
\hat p_2X^3&=&X^3(\hat p_2-\sfrac {\hbar}{R^3}(\hat x^4+\hat x^1\hat x^2)_R)~,\quad \hat p_1X^4=X^4(\hat p_1-\sfrac{\hbar}{R^4}\hat x^2_R)~.
\eea 

\subsection{Dimension 6}

In six dimensions there are many cases and we will not present all of them 
in detail. The phase spaces for each case can be reconstructed with the data 
we collect in the Appendix. Here we would like to examine in detail two 
representative examples, one step 2 and one step 5, which have properties that are absent in the 4D cases. 

\paragraph{Step 2: (0,0,0,0,13+42,14+23)~.}

We pick this representative case of step 2 nilmanifold in six dimensions 
because its momentum commutator contains a quadratic term, i.e. 
$P_{ij}^{kl}\ne 0$~, which did not happen in the 4D cases.

According to the corresponding entry in the Appendix, the basis 
1-forms are
\be
e^i=\dd x^i~, ~i=1,...,4~, \quad e^5=\dd x^5+x^3\dd x^1-x^4\dd x^2~, 
\quad e^6=\dd x^6+x^4\dd x^1+x^3\dd x^2~.
\ee
It is immediately confirmed that for the symplectic 2-form it holds that  
\be 
e^{16}+e^{25}+e^{34}=\dd x^{16}+\dd x^{25}+\dd x^{34}~,\nn
\ee 
as it is required in our analysis. 
The dual vectors are 
\be 
\theta_1=\partial_1-x^3\partial_5-x^4\partial_6~,\quad \theta_2=\partial_2+x^4\partial_5-x^3\partial_6~,\quad 
\theta_i=\partial_i~, ~i=3,\dots,6~,
\ee 
and obviously
\be 
\theta=\theta_{16}+\theta_{25}+\theta_{34}=\partial_{16}+\partial_{25}+\partial_{34}~.\nn
\ee 
Then the non-vanishing position commutators are
\be 
[\hat x^1,\hat x^6]=[\hat x^2,\hat x^5]=[\hat x^3,\hat x^4]=i~.
\ee 
The nonconstant inverse frame components read as
\be 
e^5_{\ 1}=-\hat x^3_R~,\quad e^5_{\ 2}=\hat x^4_R~,\quad e^6_{\ 1}=-\hat x^4_R~,\quad e^6_{\ 2}=-\hat x^3_R~,
\ee 
leading to the mixed commutators
\be 
[\hat x^5,\hat p_1]=[\hat x^6,\hat p_2]=-i\hbar \hat x^3_R~, \quad 
[\hat x^5,\hat p_2]=-[\hat x^6,\hat p_1]=i\hbar\hat x^4_R~.
\ee 
Eq. (\ref{momentasimp}) yields the momenta 
\bea 
\hat p_1&=&\hbar([\hat x^6,\cdot]+\hat x^3_R[\hat x^2,\cdot]+\hat x^4_R[\hat x^1,\cdot])~,\nn\\
\hat p_2&=&\hbar([\hat x^5,\cdot]-\hat x^4_R[\hat x^2,\cdot]+\hat x^3_R[\hat x^1,\cdot])~,\nn\\
\hat p_3&=&\hbar[\hat x^4,\cdot]~,\quad \hat p_4=-\hbar[\hat x^3,\cdot]~,\nn\\
\hat p_5&=&-\hbar[\hat x^2,\cdot]~,\quad \hat p_6=-\hbar[\hat x^1,\cdot]~.\nn
\eea 
Their commutation relations can be determined by direct computation and the 
non-vanishing ones are
\bea  
[\hat p_1,\hat p_3]&=&[\hat p_4,\hat p_2]=-i\hbar\hat p_5~, \nn\\
{[}\hat p_1,\hat p_4]&=&[\hat p_2,\hat p_3]=-i\hbar \hat p_6~,\nn\\
{[}\hat p_1,\hat p_2]&=&i(\hat p_5)^2+i (\hat p_6)^2~.
\eea 
We observe a quadratic commutator in the last line, which is essentially due to the non-vanishing $P_{12}^{55}=P_{12}^{66}=i$ parameters, as listed 
in the Appendix.
 
The Jacobi identities involve a nontrivial cancellation. This appears in 
the identities $\text{Jac}(\hat p_1,\hat p_2,\hat x^5)=0$ and 
$\text{Jac}(\hat p_1,\hat p_2,\hat x^5_R)=0$. Let us go through the first, since the second works the same way.
We compute
\bea 
[[\hat p_1,\hat p_2],\hat x^5]&=&[i (\hat p_5)^2+i (\hat p_6)^2,\hat x^5]
=i[(\hat p_5)^2,\hat x^5]=2\hbar\hat p_5~,\nn\\
{[}[\hat x^5,\hat p_1],\hat p_2]&=&[-i\hbar \hat x^3_R,\hat p_2]=-\hbar \hat p_5~,\nn\\
{[}[\hat p_2,\hat x^5],\hat p_1]&=&[-i\hbar\hat x^4_R,\hat p_1]=-\hbar \hat p_5~,
\eea  
where we directly computed and used that $[\hat x^3_R,\hat p_2]=[\hat x^4_R,\hat p_1]=i\hat p_5~$. Adding up the three 
terms we indeed confirm that the Jacobi identity is satisfied. 
Moreover, the rest of the nontrivial commutators involving $\hat x^a_R$ are
\bea 
[\hat x^3_R,\hat p_1]=-[\hat x^4_R,\hat p_2]=i\hat p_6~, \quad [\hat x^5_R,\hat p_1]=[\hat x^6_R,\hat p_2]=-i\hbar\hat x^3_R~,\quad
[\hat x^5_R,\hat p_2]=-[\hat x^6_R,\hat p_1]=i\hbar\hat x^4_R~.\nn
\eea 
For the compact case, the frame leads to the classical relations 
\bea 
&& x^3\to x^3+2\pi R^3 \quad \Rightarrow \quad x^5\to x^5-2\pi R^3x^1~,\quad x^6\to x^6-2\pi R^3x^2~,\nn\\
&& x^4\to x^4+2\pi R^4 \quad \Rightarrow \quad x^5\to x^5+2\pi R^4x^2~,\quad x^6\to x^6-2\pi R^4x^1~,\nn
\eea 
which imply the following off-diagonal $\tau^a_{\ i}$ components for the noncommutative case:
\be 
\tau^5_{\ 3}=\tau^6_{\ 4}=-\hat x^1_R~,\quad \tau^5_{\ 4}=-\tau^6_{\ 3}=\hat x^2_R~.
\ee 
The algebra in the compact case is easily constructed using Eqs. (\ref{compgen1}) and (\ref{compgen2}) and we do not 
write it explicitly.

\paragraph{Step 5: (0,0,12,13,14+23,15+24)~.}

The last case we highlight is a step 5 nilmanifold (the last entry of Table 10 in the Appendix),
which is the most complicated one and also the only one that leads to an 
$\hat x^a_R$-dependent quadratic term in the momentum commutator. We would like in particular to examine 
how the Jacobi identities are satisfied.

The 1-forms in this case are taken to be 
\bea 
e^1&=&\dd x^1~,\quad e^2=\dd x^2~,\quad  e^3=\dd x^3+x^2\dd x^1~,\quad e^4=\dd x^4-x^1\dd x^3~,\nn\\
e^5&=&\dd x^5+(x^4-x^1x^3)\dd x^1+(x^3+x^1x^2)\dd x^2~,\nn\\
e^6&=&\dd x^6+(x^5+2x^2x^3+\sfrac 12 x^1(x^2)^2)\dd x^1-(x^4-x^1x^3)\dd x^2+2x^1x^2\dd x^3-2x^2\dd x^4~.\nn
\eea
A direct computation shows that 
\be 
\o=e^{16}+2e^{34}-e^{25}=\dd x^{16}+2\dd x^{34}-\dd x^{25}\nn
\ee 
for the symplectic 2-form. 
Simlarly, the symplectic 2-vector satisfies
\be
\theta=\theta_{16}+\sfrac 12\theta_{34}-\theta_{25}=\partial_{16}+\sfrac 12\partial_{34}-\partial_{25}~,
\ee 
and the position commutators are 
\be 
[\hat x^1,\hat x^6]=[\hat x^5,\hat x^2]=i~,\quad [\hat x^3,\hat x^4]=\sfrac i2~.
\ee
Finding the inverse of the frame,
\bea 
e^a_{\ i}&=&\begin{pmatrix} 1 &0&0&0&0&0 \\ 0&1&0&0&0&0 \\ x^2&0&1&0&0&0 \\ 
0&0&-x^1&1&0&0 \\ x^4-x^1x^3& x^3+x^1x^2 &0&0&1&0 \\ x^5+2x^2x^3+\sfrac 12 x^1(x^2)^2&-x^4+x^1x^3&2x^1x^2&-2x^2&0&1
\end{pmatrix}^{-1}\nn\\ &=&\begin{pmatrix} 1 &0&0&0&0&0 \\ 0&1&0&0&0&0 \\ -x^2&0&1&0&0&0 \\ 
-x^1x^2&0&x^1&1&0&0 \\ -x^4+x^1x^3& -x^3-x^1x^2&0&0&1&0 \\ -x^5-2x^2x^3-\sfrac 12 x^1(x^2)^2&x^4-x^1x^3&0&2x^2&0&1
\end{pmatrix}~,
\eea
provides the mixed commutators $[\hat x^a,\hat p_i]=i\hbar e^a_{\ i}(\hat x^b_R)$~, which we do not write explicitly. 
The momenta are found to be 
\bea 
\hat p_1&=&\hbar\biggl([\hat x^6,\cdot]-2\hat x^2_R[\hat x^4,\cdot]+2\hat x^1_R\hat x^2_R[\hat x^3,\cdot]-\nn\\
&&\qquad -(\hat x^4_R-\hat x^1_R\hat x^3_R)[\hat x^2,\cdot]+(\hat x^5_R+2\hat x^2_R\hat x^3_R+\sfrac 12\hat x^1_R(\hat x^2_R)^2)[\hat x^1,\cdot]\biggl)~,\nn\\
\hat p_2&=&-\hbar\biggl([\hat x^5,\cdot]+(\hat x^3_R+\hat x^1_R\hat x^2_R)[\hat x^2,\cdot]+(\hat x^4_R-\hat x^1_R\hat x^3_R)[\hat x^1,\cdot]\biggl)~,\nn\\
\hat p_3&=&2\hbar([\hat x^4,\cdot,]-\hat x^1_R[\hat x^3,\cdot])~,\nn\\
\hat p_4&=&-2\hbar([\hat x^3,\cdot]+\hat x^2_R[\hat x^1,\cdot])~,\nn\\
\hat p_5&=&\hbar [\hat x^2,\cdot]~,\quad \hat p_6=-\hbar[\hat x^1,\cdot]~,\nn
\eea
with quadratic commutation relations (we do not explicitly write the ones that are simply linear):
\bea \label{p12s53}
[\hat p_1,\hat p_2]&=&-i\hbar\hat p_3+\sfrac i2(\hat p_5)^2+i\hat x^2_R(\hat p_6)^2-i\hat x^1_R\hat p_5\hat p_6~,\\
{[}\hat p_1,\hat p_4]&=&-i\hbar\hat p_5+2i(\hat p_6)^2~.
\eea 
We observe that the commutator (\ref{p12s53}) contains $\hat x^a_R$-dependent quadratic terms. 
In the present case, the Jacobi identities involve highly nontrivial cancellations. For example,
$
\text{Jac}(\hat p_1,\hat p_2,\hat x^5_R)
$ contains the terms 
\bea 
[[\hat p_1,\hat p_2],\hat x^5_R]&=&\hbar \hat p_5-(\hat p_6)^2-\hbar \hat x^1_R\hat p_6~,\nn\\
{[}[\hat x^5_R,\hat p_1],\hat p_2]&=&-\sfrac 32\hbar\hat p_5+(\hat p_6)^2-\hbar \hat x^1_R\hat p_6~,\nn\\
{[}[\hat p_2,\hat x^5_R],\hat p_1]&=&\sfrac 12 \hbar \hat p_5+2\hbar \hat x^1_R\hat p_6~,
\eea
and we observe that they sum to zero. A number of such cancellations occurs for the rest of the Jacobi identities too.

The treatment of the compact case follows the same lines as in the previous examples and therefore we do not present it explicitly.

\section{Discussion}

The main arena of quantum mechanics is phase space, which is quantized and without points in the classical sense of geometry, due to the 
uncertainty principle. On the other hand, general relativity accounts for the gravitational interaction by describing the dynamics of 
spacetime. The question of how these two theories become compatible is the most challenging conceptual problem in theoretical 
physics today. It is conceivable that one way that might illustrate the path towards quantum gravitational physics is to 
determine a framework where phase space and dynamical spacetime are reconciled in a dynamical theory of phase space (see e.g. 
Ref. \cite{Freidel:2014qna} for a recent argumentation).

In this paper we employed an algebraic point of view and examined the algebraic properties of noncommutative phase spaces 
in the presence of a nontrivial frame. Although we did not study any dynamics, our results indicate that this can be possible at 
a later stage. In particular we showed that there exist consistent algebraic structures that incorporate quantized phase 
spaces with curvature and we studied a particular class of explicit examples. The new element that did not appear in previous 
similar approaches is that two copies of the noncommutative algebra are necessary, consisting of operators 
with a left and right action respectively. Most importantly, in curved cases these operators play an asymmetric role and 
satisfy different commutation relations with the momenta, a fact that is hidden in the flat case. This result should be taken into account in any attempt to construct a dynamical theory of phase space.

Considering that we are working in the framework of noncommutative geometry, one of the most motivating ways to think of an 
incarnation of such a dynamical theory of phase space is matrix models. Recall that in the well-known IKKT model \cite{Ishibashi:1996xs}, 
spacetime emerges dynamically (see e.g. Ref. \cite{Nishimura:2012xs} for a recent review). Moreover, the model can be 
quantized via an integral over matrices, given by the partition function
\be 
\mc Z=\int \dd A\dd \Psi e^{-S}~, \quad S=-\sfrac 1{4}\text{Tr}~[A_M,A_N]^2+S_{\text{matter}}~,
\ee 
with $A_M$ being ten Hermitian matrices. Correlation functions may be similarly defined and they are in principle 
computable with analytical or numerical methods. Classical solutions of the model are typically noncommutative spaces, where 
coordinate operators are identified with the matrices. On the other hand, according to the general arguments that we presented 
here, it would be preferable to obtain classical solutions that correspond to noncommutative phase spaces, such as the ones 
studied in this paper, in order to understand the role of the gravitational field too.
Presumably, these are solutions of an extended matrix model that can account for the 
dynamics of phase space and can be quantized in a way similar to the above. We hope to report on this in a future publication.

In this paper we started with four basic assumptions, namely
\begin{itemize}
 \item parallelizability, or equivalently existence of a globally well defined frame,
 \item symplectic structure,
 \item Leibniz rule and 
 \item Jacobi identities.
\end{itemize}
Then we set the position commutators equal to the 
components of the symplectic 2-vector and implementing the frame by setting the mixed commutator between positions 
and momenta proportional to it. Consistency of these relations allowed us to determine the general form of the 
momentum operators as well as their commutation relation, which turned out to be quadratic in the momenta in 
accord with previous results \cite{Madorebook}. 

In the process of our investigation we emphasized the distinct role of left and right acting operators and discussed the 
symplectic duality among the two sets. Although in simple cases, like the d-plane or the d-torus, this does not have any nontrivial 
consequences, departure from flatness breaks the symmetric role among the two. In particular, consistency of the formalism led us 
to associate the noncommutative frame to the set of right acting operators, when the observables of the theory lie in the pool 
of left acting ones. This made it necessary to consider an extended algebra of position operators $\hat x^a_L$~, momentum operators 
$\hat p_i$ and quantized coordinate operators $\hat x^a_R$. This extended algebra was fully determined and it turned out that all the Jacobi identities are satisfied.

The general approach finds an elegant realization in a class of spaces which are known as nilmanifolds. These are iterated
nontrivial fibrations of tori over tori and they yield several cases of symplectic manifolds in four and six dimensions. 
These symplectic cases were classified already in Ref. \cite{goze} and here we reclassified them according to their nilpotency step. 
Then we applied the general results and discussed some benchmark cases in detail. 

In analogy to the compactification of a d-plane to a d-torus, manifolds based on nilpotent Lie algebras can be also compactified 
with a similar procedure based on identification conditions. It is well known that in compact cases, the position operators 
have to be exponentiated in order to be single valued. This holds true in the case of nilmanifolds too, although 
one has to be cautious about some additional complications due to the liberation between left and right operators. 
In particular it turned out that the identification conditions have to be imposed on the right operators, thus 
compactifying the noncommutative manifold, while the left acting operators are simply exponentiated, thus rendering 
the positions single valued. Similarly to the flat case, where two dual tori appear in the compactification process, 
in the curved case we encounter two dual nilmanifolds. 

We already emphasized that the main goal would be to derive some dynamical theory of phase space that would be relevant 
for quantum gravity. Apart from this, there are four more immediate and clear paths that call for further investigation. 
First, an important next step of the present analysis is to define and 
compute the curvature of the quantized spaces in question. One can 
expect to derive an expression that converges to the classical curvature in 
the commutative limit but it carries more terms in the quantum case. 
A similar approach was employed in Refs. \cite{Buric:2006di,Buric:2007hb,Buric:2011dd}. 
Such a task will also assist in understanding the common features and the differences with other 
recent approaches, such as \cite{Steinacker:2010rh,Blaschke:2010qj,Blaschke:2010rg,Aschieri:2009ky,Aschieri:2012vc,Yang:2013aia}.
Second, a different direction would be to go beyond the symplectic case. 
In general, nilmanifolds are not always symplectic, while symplectic nilmanifolds are not always \textit{only} symplectic. 
At least in six dimensions, 
all nilmanifolds admit generalized complex structures, as proven in Ref. 
\cite{sixmanifolds}. This fact was used in the analysis of 
Dirac structures on step 2 nilmanifolds in Ref. \cite{Chatzistavrakidis:2013wra}. The task would be to study the quantization of these structures 
as well. The third direction is to consider the quantization in the presence of sources. Although this is known for planes and tori, it is less obvious 
how straightforward it will be to implement sources for the nilmanifolds 
in the quantum case. A final possibility would be to go beyond nilmanifolds 
and examine what other spaces could be handled with the techniques of 
the present work. For example an obvious challenge is to examine symplectic solvmanifolds, which are more complicated that nilmanifolds.

\paragraph{Acknowledgements.} The author would like to thank M. Buri\'c, F.F. Gautason, L. Jonke, O. Lechtenfeld, J. Madore, P. Schupp 
and R. Szabo for discussions related to the content of the paper.  

\vspace{30pt}
\appendix
\textbf{\Large{Appendix}}
\section{Additional data for symplectic nilmanifolds}

In this Appendix we collect all the necessary additional data for the 
construction of the noncommutative phase space of symplectic nilmanifolds. 
We recall that this is given as
\bea 
[\hat x^a,\hat x^b]&=&i\theta^{ab}~,\nn\\
{[}\hat x^a,\hat p_i]&=&i\hbar e^a_{\ i}~,\nn\\
{[}\hat p_i,\hat p_j]&=&M_{ij}+N_{ij}^{\ k}\hat p_k+P_{ij}^{kl}\hat p_k\hat p_l~.\nn
\eea 
The parameters $\theta^{ab}$ can be read off from the Tables 1-5 in the main text. They are antisymmetric 
and their value is $\pm 1$ along the directions of 
the symplectic structure. The parameters $e^a_{\ i}$ can be read off from the Tables 
6-10 that we present in this Appendix, in the column titled ``Frame: $e^i-\d^i_{\ a}\dd x^a$''~. The frame is written as a d-tuple 
$(\d e^1_{\ a}(x)\dd x^a,\d e^2_{\ a}(x)\dd x^a,...,\d e^{\text{d}}_{\ a}(x)\dd x^a)~$, where $\d e^i_{\ a}=e^i_{\ a}-\d^i_{\ a}~$.
The parameters $M_{ij}$ are always vanishing, so they are not presented. 
Moreover, we have checked that the parameters that determine the linear term in the momentum commutator are 
generically given as $$N_{ij}^{\ k}=-i\hbar f^k_{\ ij}~.$$ 
This was proven analytically for step 2 nilmanifolds in Section 4.2, but it turns out that it holds for any step.
Therefore it is unnecessary to present these parameters in the tables. 
 Finally, we present the quadratic parameters 
$P_{ij}^{kl}$ and we remind that they are antisymmetric in the lower indices and symmetric in the upper ones. Note that the entries that 
do not satisfy the proposition of Section 4.2 are not examined and they are marked as ``$N/A$''.
 
\renewcommand{\arraystretch}{1.5}
\begin{center}
\footnotesize{\boxed{
\begin{tabular}{cccc}
 \underline{Class} & \underline{Step} & 
\underline{Frame: $e^i-\d^i_{\ a}\dd x^a$}  & \underline{$P^{kl}_{ij}$} 
 \\[4pt]
(0,0,0,12)  & 2 & $(0,0,0,x^2\dd x^1)$&0\\[4pt]
(0,0,42,12) & 3 &  $(0,0,-(x^4+x^1x^2)\dd x^2,x^2\dd x^1)$&0
\end{tabular}}\captionof{table}{Data for symplectic nilmanifolds in 4D.}}
\end{center}

\renewcommand{\arraystretch}{1.5}

\begin{center}
\footnotesize{\boxed{
\begin{tabular}{ccc}
 \underline{Class}  & \underline{Frame: $e^i-\d^i_{\ a}\dd x^a$} & \underline{$P^{kl}_{ij}$} 
 \\[4pt]
(0,0,0,0,0,12) & $(0,0,0,0,0,x^2\dd x^1)$&0\\[4pt]
(0,0,0,0,13+42,14+23)  &  $(0,0,0,0,x^3\dd x^1-x^4\dd x^2,x^4\dd x^1+x^3\dd x^2)$&  $P_{12}^{55}=P_{12}^{66}=i$ \\[4pt]
(0,0,0,0,12,13)  &  $(0,0,0,0,-x^1\dd x^2,x^3\dd x^1)$&0\\[4pt]
(0,0,0,0,12,34)  & $(0,0,0,0,x^2\dd x^1,x^4\dd x^3)$&$P^{56}_{13}=-i/2$\\[4pt]
(0,0,0,0,12,14+23) & $(0,0,0,0,-x^1\dd x^2,x^3\dd x^2-x^1\dd x^4)$&
$P^{66}_{24}=-i$\\[4pt]
(0,0,0,12,13,23)  & $(0,0,0,-x^1\dd x^2,x^3\dd x^1,-x^2\dd x^3)$&0
\end{tabular}}\captionof{table}{Data for step 2 symplectic nilmanifolds in 6D.}}
\end{center}

\begin{center}
\footnotesize{
\boxed{
\begin{tabular}{ccc}
 \underline{Class}  &  \underline{Frame: $e^i-\d^i_{\ a}\dd x^a$} & \underline{$P^{kl}_{ij}$} 
 \\[4pt]
(0,0,0,0,12,14+25)  & $(0,0,0,0,-x^1\dd x^2,x^5\dd x^2-x^1\dd x^4)$&0\\[4pt]
(0,0,0,0,12,15)  &  $(0,0,0,0,-x^1\dd x^2,(x^5-x^1x^2)\dd x^1)$
&0 \\[4pt]
(0,0,0,12,14+23,13+42)  & \begin{tabular}{c} $(0,0,0,-x^1\dd x^2,(x^4-x^1x^2)\dd x^1+x^3\dd x^2,$\\ $\sfrac 12 x^3\dd x^1-x^4\dd x^2-\sfrac 12 x^1\dd x^3)$ \end{tabular}&$P_{12}^{66}=\sfrac i2$\\[4pt]
(0,0,0,12,14,13+42)  & $(0,0,0,-x^1\dd x^2,(x^4-x^1x^2)\dd x^1,-x^4\dd x^2-x^1\dd x^3)$&0\\[4pt]
(0,0,0,12,14,23+24)  & $(0,0,0,x^2\dd x^1,x^4\dd x^1,-\sfrac 12(x^2)^2\dd x^1+x^4\dd x^2-x^2\dd x^3)$&0\\[4pt]
(0,0,0,12,13,14)  & $(0,0,0,-x^1\dd x^2,-x^1\dd x^3,(x^4-x^1x^2)\dd x^1)$&0\\[4pt]
(0,0,0,12,13,24) & $(0,0,0,x^2\dd x^1,-x^1\dd x^3,(x^4+x^1x^2)\dd x^2)$&$P^{56}_{23}=-\sfrac i2$\\[4pt]
(0,0,0,12,13,14+23)  &\begin{tabular}{c} $(0,0,0,\sfrac 12 x^2\dd x^1-\sfrac 12 x^1\dd x^2,-x^1\dd x^3,$\\$(x^4-\sfrac 12 x^1x^2)\dd x^1-x^2\dd x^3)$\end{tabular}&0
\end{tabular}}\captionof{table}{Data for step 3 symplectic nilmanifolds in 6D.}}
\end{center}

\begin{center}
\footnotesize{
\boxed{
\begin{tabular}{ccc}
 \underline{Class}  &  \underline{Frame: $e^i-\d^i_{\ a}\dd x^a$} & \underline{$P^{kl}_{ij}$} 
 \\[4pt]
(0,0,0,12,14-23,15+34)  & \begin{tabular}{c}$(0,0,0,-x^1\dd x^2,(x^4-x^1x^2)\dd x^1+x^2\dd x^3,$\\$x^5\dd x^1+(x^4-x^1x^2)\dd x^3)$\end{tabular}&$P^{55}_{13}=i$\\[4pt]
(0,0,0,12,14,15)  & \begin{tabular}{c} $(0,0,0,-x^1\dd x^2,\sfrac 12 (x^1)^2\dd x^2-x^1\dd x^4,$\\$-x^1\dd x^5+\sfrac 12 (x^1)^2\dd x^4-\sfrac 16(x^1)^3\dd x^2)$\end{tabular} &0\\[4pt]
(0,0,0,12,14,15+24) &  \begin{tabular}{c}$(0,0,0,-x^1\dd x^2,\sfrac 12 (x^1)^2\dd x^2-x^1\dd x^4,$\\$x^4\dd x^2-x^1\dd x^5+\sfrac 12 (x^1)^2\dd x^4-\sfrac 16(x^1)^3\dd x^2)$\end{tabular}&0\\[4pt]
(0,0,0,12,14,15+23+24) &  \begin{tabular}{c}$(0,0,0,-x^1\dd x^2,\sfrac 12 (x^1)^2\dd x^2-x^1\dd x^4,$\\$(x^3+x^4)\dd x^2-x^1\dd x^5+\sfrac 12 (x^1)^2\dd x^4-\sfrac 16(x^1)^3\dd x^2)$\end{tabular}&\begin{tabular}{c}
$P^{56}_{24}=-\sfrac i2$\\$P^{66}_{25}=-i$
\end{tabular}\\[4pt]
(0,0,0,12,14,23+15) &\begin{tabular}{c}  $(0,0,0,-x^1\dd x^2,\sfrac 12 (x^1)^2\dd x^2-x^1\dd x^4,$\\$x^3\dd x^2-x^1\dd x^5+\sfrac 12 (x^1)^2\dd x^4-\sfrac 16(x^1)^3\dd x^2)$\end{tabular}&\begin{tabular}{c}
$P^{56}_{24}=-\sfrac i2$\\$P^{66}_{25}=-i$
\end{tabular}\\[4pt]
(0,0,12,13,23,14) &  $N/A$\\[4pt]
(0,0,12,13,23,14-25) &  $N/A$\\[4pt]
(0,0,12,13,23,14+25) &  $N/A$
\end{tabular}}\captionof{table}{Data for step 4 symplectic nilmanifolds in 6D.}}
\end{center}

\begin{center}
%\begin{sidewaystable}
\footnotesize{
\boxed{
\begin{tabular}{ccc}
 \underline{Class}  &  \underline{Frame: $e^i-\d^i_{\ a}\dd x^a$} 
 & \underline{$P^{kl}_{ij}$} 
 \\[4pt]
(0,0,12,13,14,15) &\begin{tabular}{c} $(0,0,-x^1\dd x^2,-x^1\dd x^3+\sfrac 12 (x^1)^2\dd x^2,$\\$-x^1\dd x^4+\sfrac 12 (x^1)^2\dd x^3-\sfrac 16(x^1)^3\dd x^2,$\\
$(x^5-x^1x^4+\sfrac 12(x^1)^2x^3-\sfrac 16(x^1)^3x^2)\dd x^1)$\end{tabular}&0\\[4pt]
(0,0,12,13,14,15+23) & $N/A$\\[4pt]
(0,0,12,13,14+23,15+24)  & \begin{tabular}{c}$(0,0,x^2\dd x^1,-x^1\dd x^3,$\\$(x^4-x^1x^3)\dd x^1+(x^3+x^1x^2)\dd x^2,$\\$(x^5+2x^2x^3+\sfrac 12 x^1(x^2)^2)\dd x^1$\\
$-(x^4-x^1x^3)\dd x^2$\\$+2x^1x^2\dd x^3-2x^2\dd x^4)$\end{tabular}&\begin{tabular}{c}
$P^{55}_{12}=\sfrac i2$\\$ P^{66}_{14}=2i$\\
$P^{66}_{12}= i\hat x^2_R$\\$P^{56}_{12}=-\sfrac i2\hat x^1_R$\end{tabular} 
\end{tabular}}\captionof{table}{Data for step 5 symplectic nilmanifolds in 6D.}}
%\end{sidewaystable}
\end{center}

\end{document}